%% file: main_arxiv.tex
\documentclass[10pt]{article}
\usepackage{arxiv}
\usepackage[utf8]{inputenc} 
\usepackage[T1]{fontenc}  
\usepackage{mathpazo}
\usepackage{amsmath}
\usepackage{eulervm} 
\usepackage{bm}
\usepackage{graphicx}
\usepackage{import}
\usepackage{caption}
\usepackage{subcaption}
\usepackage{cool}
\usepackage[bottom]{footmisc}
\usepackage{titlesec}
\usepackage{enumitem}
\usepackage[round]{natbib}
\usepackage{hyperref}
\usepackage[capitalise]{cleveref}
\usepackage{tikz}
\usepackage{pgfplots}
\pgfplotsset{compat=newest}
\titleformat*{\subsubsection}{\normalsize\itshape}
\usepackage{rotating}
\usepackage{array}

\newcommand{\av}[1]{\langle#1\rangle}

\DeclareMathAlphabet{\mathsfbf}{OT1}{\sfdefault}{bx}{n}
\DeclareMathOperator*{\argmin}{arg\,min}

\title{Machine learning-based multiscale constitutive modelling: Development and application to dual-porosity mass transfer}

\author{
  Mark Ashworth\\
  Institute of GeoEnergy Engineering\\
  Heriot-Watt Unversity, Edinburgh \\
  \texttt{ma174@hw.ac.uk} \\
   \And
  Ahmed Elsheikh \\
  Institute of GeoEnergy Engineering\\
  Heriot-Watt Unversity, Edinburgh \\
  \texttt{a.elsheikh@hw.ac.uk} \\
   \And
  Florian Doster\\
  Institute of GeoEnergy Engineering\\
  Heriot-Watt Unversity, Edinburgh \\
  \texttt{f.doster@hw.ac.uk} \\
}

\date{}

\begin{document}
\newlength\figureheight
\newlength\figurewidth
\maketitle
\captionsetup[figure]{labelfont={bf, small},labelformat={default},labelsep=period,name={Figure}, font=small}
\begin{abstract}
In multiscale modelling, multiple models are used simultaneously to describe scale-dependent phenomena in a system of interest. Here we introduce a machine learning (ML)-based multiscale modelling framework for modelling hierarchical multiscale problems. In these problems, closure relations are required for the macroscopic problem in the form of constitutive relations. However, forming explicit closures for nonlinear and hysteretic processes remains challenging. Instead, we provide a framework for learning constitutive mappings given microscale data generated according to micro and macro transitions governed by two-scale homogenisation rules. The resulting data-driven model is then coupled to a macroscale simulator leading to a hybrid ML-physics-based modelling approach. Accordingly, we apply the multiscale framework within the context of transient phenomena in dual-porosity geomaterials. In these materials, the inter-porosity flow is a complex time-dependent function making its adoption within flow simulators challenging. We explore nonlinear feedforward autoregressive ML strategies for the constitutive modelling of this sequential problem.  We demonstrate how to inject the resulting surrogate constitutive model into a simulator. We then compare the resulting hybrid approach to traditional dual-porosity and microscale models on a variety of tests. We show the hybrid approach to give high-quality results with respect to explicit microscale simulations without the computational burden of the latter. Lastly, the steps provided by the multiscale framework herein are sufficiently general to be applied to a variety of multiscale settings, using different data generation and learning techniques accordingly.
\end{abstract}

\section{Introduction}
When modelling we are often posed with the dilemma of choosing between models of varying levels of accuracy, complexity and efficiency. Such problems are exemplified in the subsurface. In these settings, strong heterogeneities exist within the rock fabric leading to multiscale phenomena. One solution is to model at the heterogeneity level (microscale), capturing the full range of scale-dependent phenomena with appropriate post-processing. However, at large scales, computational and information requirements preclude the use of such detailed simulation models. Instead, in situations where there is weak coupling between scales, a so-called separation of scales exists and we can use coarse representations. Accordingly, we have a hierarchy of models, where microscale information is incorporated implicitly within macroscale models in the form of constitutive relations. However, often these constitutive models are based on empiricism using simple considerations such as linearity (\citealt{Weinan2011}). Whilst practical, use of macroscopic models with simplified constitutive relations can lead to undesirable inaccuracies. This deficiency motivates the use of multiscale modelling approaches. That is, to combine the detail of microscale models with the efficiency of macroscale models.  

Here we are concerned with the use of multiscale methods with application to hierarchical multiscale problems. Generally, multiscale methods can be classified as analytical or computational (\citealt{Weinan2011}), although significant overlap may exist between the two. In situations where analytical methods are impractical or intractable, it is desirable instead to use computational approaches. One such computational approach, informed by developments in the latter part of the twentieth century, is computational homogenisation (\citealt{Feyel1999, Kouznetsova2001, Miehe2002}). In this approach, microscopic boundary value problems are nested within points in the macroscopic domain, with the solution of the former providing information for the latter. Notably, computational homogenisation has emerged as a powerful tool for modelling multiscale problems as diverse coupled multiphysics (\citealt{Ozdemir2008b, Su2011, Kaessmair2018}) and transient diffusion problems (\citealt{Larsson2010, Ramos2017, Waseem2020}) to name but a few. Nonetheless, whilst accurate, these approaches still remain expensive to use in practice (\citealt{Wang2018}). This expense motivates the use of so-called sequential methods. Here calculations are typically made offline to create a look-up table. The look-up table is then called online, using an appropriate interpolater where necessary. The downsides of the look-up table approach are we acquire little knowledge about the function, the quality of our dataset and how well we may generalise to samples not included in the table. Alternatively, we can \textit{learn} a surrogate model offline, given our input-output data and other possible information. Learning such mappings is one of the fundamental goals of machine learning (ML). 

Machine learning approaches to generating surrogate constitutive models can be traced back several decades. For example, in the pioneering work of \cite{Ghaboussi1991} the authors used so-called neural networks to represent the constitutive behaviour of concrete under various loading conditions. From this and other similar early works (\citealt{Goh1995, Ghaboussi1998}), users have incorporated data-driven surrogate models within multiscale modelling strategies for a variety of applications (\citealt{Unger2009, Hambli2011, Asproulis2013}). More recently, work has been done for modelling time-dependent constitutive information using so-called recurrent neural networks (\citealt{Wang2018, Ghavamian2019, Chen2021}). Despite the success of these recurrent algorithms for sequential modelling, they can be challenging to implement and to understand (\citealt{Pascanu2013, Bai2018}). Further, \citealt{Bai2018} showed more recently autoregressive feedforward neural network architectures that give superior performance compared to recurrent structures for a wide variety of sequence modelling tasks. Given the variety of multiscale problems and machine learning methods, it would be useful to have a framework detailing the key questions and considerations for the generation and use of surrogate constitutive models within sequential multiscale approaches. 

The aim and novelty of this work is to introduce and apply a machine learning-based multiscale framework for modelling complex constitutive behaviours such as those due to nonlinearity and hysteresis. Our framework is broken down into four key components, with the starting point being a homogenisation procedure. Here we use a two-scale homogenisation approach such as those forming the basis for computational homogenisation methodologies. In this framework, the homogenisation procedure provides the means for generating consistent data to learn from. Further, the procedure also allows insight into the quantities involved in the constitutive mapping. With the scale transition rules from the homogenisation procedure in place, the remaining parts of the framework involve: Generating the data, learning the model and coupling the result to a physics-based simulator. 

For our application we consider the diffusive mass transfer involved in dual-porosity (or dual-continuum more generally) flow modelling. In the dual-porosity (DP) setting, a strong contrast in conductivities exists between the different porosity levels. A classical example of such a material is naturally fractured rock. Here, the matrix has significantly lower permeability than the fractures. As a result, the matrix introduces a history-dependent memory effect associated with the mass exchange between the two porosity levels. Accordingly, in the presence of fracture dynamics the inter-porosity exchange is non-local in time, requiring a convolutional product for its description. Further, even for cases involving static fracture pressures, challenges still remain in describing inter-porosity exchange. For these latter cases, and for simple geometries, analytical solutions for the inter-porosity exchange term are possible. However, solutions for these static fracture pressure cases are still explicit in time, and are formed from a complex infinite series. Accordingly, describing inter-porosity flow for both static and dynamic fracture behaviours remains difficult with traditional simulation methods. These challenges necessitate simpler descriptions for inter-porosity exchange, leading to the commonly used linear description involving involving the difference between averaged pressures (\citealt{Warren1963}). However, this simplification is known to lead to measurable errors in flow behaviour, particularly at early-time (\citealt{Zimmerman1993, Ashworth2020}). Instead of making any such assumptions, we use a data-driven constitutive model learnt on time series data coming directly from the microscale. In line with the aims of this work to introduce and apply our framework, we consider only cases of static fracture pressure. Accordingly, we can give an overview of the framework on a problem without significant complexity, using easy-to-generate data and established learning algorithms. We subsequently compare various nonlinear feedforward autoregressive algorithms approaches to model the time-dependent problem. The successful model is then coupled to a dual-porosity simulator in the \textsc{Matlab} Reservoir Simulation Toolbox (MRST) (\citealt{Lie2012, Lie2019}).

We structure the remainder of this paper as follows: In Section 2, we introduce the framework, detailing the key elements and considerations therein. In Section 3, we start the application of the framework through the two-scale homogenisation approach to derive the macroscopic dual-porosity model. In Section 4, we introduce a dual-porosity model problem that will serve as the testbed for the remaining parts of the multiscale framework. Further, given this model problem we derive an often-used linear transfer model. In Section 5, we generate the dataset for learning. To do so, we use analytical solutions for the given model problem, whilst also considering the end use of the ML model in a numerical simulator. In Section 6, we learn our surrogate constitutive model, comparing a variety of learning algorithms for modelling the time-dependent problem. In Section 7, we couple the resulting ML model to a physics-based simulator, leading to a hybrid ML-physics approach. We then test the resulting hybrid approach against a dual-porosity model equipped with the traditionally used linear transfer model and microscale simulations. In Section 8, we finish with conclusions and recommendations for future work. Lastly, for reproducibility, the codes used in this work are available as open-source \footnote{\href{https://github.com/mashworth11/ML-MM}{{\fontfamily{qcr}\selectfont https://github.com/mashworth11/ML-MM}}}. 

\section{Machine learning-based constitutive modelling framework}
\begin{figure}[h]
\centering
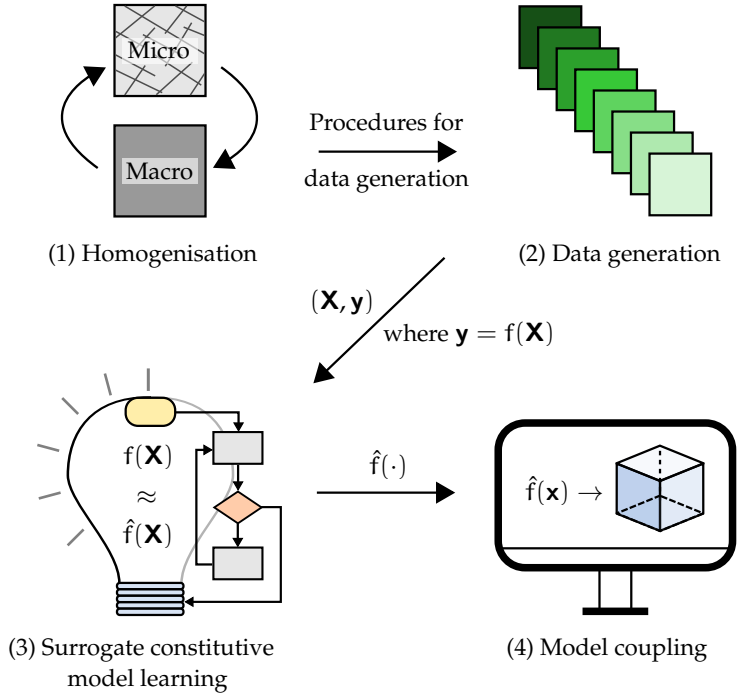
\caption{The machine learning-based multiscale constitutive modelling framework.}
\label{fig:1}
\end{figure}

Here we introduce the ML-based multiscale constitutive modelling framework and the key concepts therein. The framework itself can be broken into four key components: Homogenisation, data generation, surrogate constitutive model learning, and ML and physics model coupling (\cref{fig:1}). However, we stress that these components are interrelated and should be understood in parallel as opposed to isolated procedures.  

\subsection{Homogenisation}
Homogenisation is concerned with the question: How do we link between micro and macroscales, providing constitutive information from the former to the latter? Various approaches exist for linking between scales such as asymptotic homogenisation (\citealt{Sanchez1980, Auriault2010}), the method of volume averaging (\citealt{Quintard1993, Whitaker1999}), and computational homogenisation approaches (\citealt{Feyel1999, Kouznetsova2001}). Depending on underlying requirements (e.g. a-priori knowledge for scaling parameters in the case of asymptotic approaches) and resources, some upscaling strategies may be more suitable than others. Regardless, within the context of this framework, the homogenisation procedure provides the guide for generating consistent \textit{offline} data for training and validating a machine learning model. We note, one could consider the homogenisation step as part of the data generation process. However, we choose to distinguish these two steps within the current framework to highlight their respective importance and considerations therein. 

\subsection{Data generation}
Data generation is concerned with the question: How will we generate sample data for our learning algorithm? Microscale results can be expensive when numerically derived due to limited computational resources and/or high problem dimensionality. Managing this trade-off to ensure the dataset (sample set) is as informative as possible is referred to as experimental design (or design of experiments) (\citealt{Razavi2012}). Traditional approaches to experimental design involve the use of various sampling strategies (e.g. Monte Carlo, Latin hypercube, quasi-Monte Carlo methods) to build the dataset prior to learning. However, these so-called one-shot strategies can either contain too few samples for an accurate surrogate (undersampling) or more samples than we need (oversampling), thus wasting precious resources (\citealt{Xiao2018}). An alternative to one-shot approaches is to use feedback from the ML training process to find samples that provide the maximum information gain when labelled and used to update the model. These so-called active learning (or sequential sampling) approaches can address the under/oversampling trade-off described previously (\citealt{Crombecq2011b}). 

In addition to data-driven ML approaches, recent advances have seen the development of theory-driven ML approaches. Here the idea is to use physics such as partial differential equations, initial and boundary condition information, symmetry or invariance properties, and thermodynamical considerations (\citealt{Ling2016, Wang2018, Raissi2019, Zhu2019, Masi2021}), to constrain the learning process. Accordingly, we can supplement missing or difficult to obtain data. Further, theory-driven approaches also act to regularise the learning process and ensure the physical consistency of our model. However, in this work we consider purely data-driven methods and save the incorporation of theory into the learning process for future studies.  

\subsection{Surrogate constitutive model learning}
Surrogate constitutive model learning is concerned with the question: Given the properties of our data, the type of prediction problem and the end use for the resulting ML model, what is the most appropriate learning strategy to use? Training ML algorithms can be expensive from both data and computational points of view. If our sample set is poor, complex algorithms run the risk of failing to generalise when given new data (overfitting). Conversely, if our model is too simple we risk failing to extract meaningful relationships from our data (underfitting). Besides addressing underfitting and overfitting, we may also want algorithms that are intrinsically interpretable and quick to retrain in light of new data (e.g. linear regression, tree-based methods); give us an insight into aleatoric (data) and/or epistemic (model) uncertainties (e.g. Gaussian processes, Bayesian learning); and provide an embedded approach to feature (input) engineering in light of complex and/or high dimensional data (e.g. deep learning). Ultimately, within the context of multiscale modelling, the choice of learning method should be guided by considering the interdependence between data quality and availability, ML model accuracy and practicality, and the model's end use in physics-based simulation. 

\subsection{Model coupling}
Model coupling is concerned with the question: How will we use our learnt constitutive model in our macroscale physics simulator? Ideally, the ML model can be implemented within a numerical solver without significant intrusion. Accordingly, important considerations at this stage may be the accessibility of data during simulation to pass to the ML model; the ML model inference time; how predictions will be fed into the numerical model; and in the case of nonlinear materials, how we extract state-dependent material properties such as permeability and stiffness from our surrogate constitutive model. These latter points will depend on whether the constitutive model is to be used in an explicit or implicit manner. We describe the former, as those predictions made using previously observed and converged quantities. Subsequently, in a nonlinear solver the resulting surrogate constitutive model will only appear in the residual. Alternatively, if the prediction also depends on quantities at the current iteration level, parameters from the surrogate associated with these quantities will be needed in the Jacobian. For example, in \cite{Hashash2004} the authors consider nonlinear deformation problems and derive a closed-form expression for calculating the tangent stiffness tensor from a neural network. Accordingly, the authors provide consistent Jacobian entries for use in a Newton-Raphson scheme. For this work, our surrogate will be explicitly coupled due to the nature of the constitutive relation being injected. Nonetheless, the example above highlights the type of considerations we need to make in order to embed ML models within traditional physics-based simulations. 

\section{Application of the framework: two-scale homogenisation for dual-porosity flow problems}
With the details of the framework in-hand we now progress to its application. For our application we consider problems associated with dual-porosity flow, such as those observed in multiscale geomaterials. Accordingly, in this section we derive the dual-porosity flow model by way of a two-scale homogenisation approach developed for transient diffusion problems (\citealt{Larsson2010, Ramos2017, Waseem2020}). 

\subsection{Preliminaries}
The homogenisation procedure relies on the definition of representative elementary volumes $\Omega$ (REV) taken from a macroscopic domain $\overline{\Omega}$. Accordingly, we are interested in quantities varying at two distinct scales: Those variations at points within an REV $\bm{x}\in\Omega$, and those variations at material points within the macroscopic body $\overline{\bm{x}}\in\overline{\Omega}$. Note, quantities written with $\overline{\phantom{x}}$ denote macroscopic quantities. 

The REV is defined according to a separation of scales in terms of geometrical information and the wavelengths of physical processes involved. Formally, the separation of scales is defined as 
\begin{equation}
    \ell \ll l \ll L, \label{eqn:1}
\end{equation}
where $\ell$, $l$ and $L$ denote the characteristic lengths of the heterogeneities, REV and macroscopic body respectively. 

Our reference material is naturally fractured rock. Accordingly, the material is composed of a low permeability matrix (denoted by subscript 1) and a high permeability fracture material (denoted by subscript 2). We assume homogeneity and isotropy of each material as well as isotropic geometrical distributions at the microscale.

Next, we introduce the following volume averaging relations
\begin{equation}
    \av{z} = \frac{1}{|\Omega|}\int_\Omega z \text{ }dV, \quad \av{z}_\alpha = \frac{1}{|\Omega_\alpha|}\int_{\Omega_\alpha} z \text{ } dV, \label{eqn:2}
\end{equation}
where $z$ is an arbitrary field quantity. Accordingly, $\av{\cdot}$ denotes the volume average over the whole REV domain, whilst $\av{\cdot}_\alpha$ denotes the volume average over the constituent volume $\Omega_\alpha\subset\Omega$. The two volume averages defined in \cref{eqn:2} are related such that
\begin{equation}
    \av{z} = v_1\av{z}_1 + v_2\av{z}_2, \label{eqn:3}
\end{equation}
where $v_\alpha=|\Omega_\alpha|/|\Omega|$ is the volume fraction of phase $\alpha$. 

Lastly, we define the following relations between rates of change and their averages, such that 
\begin{equation}
    \av{\dot{z}} = \Bigl<\pderiv{z}{t}\Bigr> = \pderiv{}{t}\av{z} = \dot{\av{z}}, \label{eqn:4}
\end{equation}

\subsection{Mass balance at the macro and microscales}
The dual-porosity model is a multi-continuum model at the macroscale described by the continuum-wise mass balance as
\begin{align}
        \dot{\overline{m}}_1 + \nabla\cdot\overline{\bm{w}}_1 &= \overline{r}_1, \label{eqn:5} \\
        \dot{\overline{m}}_2 + \nabla\cdot\overline{\bm{w}}_2 &= \overline{r}_2, \label{eqn:6}
\end{align}
where $\dot{\overline{m}}_\alpha$, $\overline{\bm{w}}_\alpha$ and $\overline{r}_\alpha$ denote the rate of macroscopic mass content change, mass flux vector, and inter-continuum rate of mass transfer for continuum $\alpha$ respectively. Further, mass conservation requires $\overline{r}_2=-\overline{r}_1$. 

In addition to appropriate boundary and initial conditions, we require constitutive relations for the macroscopic quantities $\overline{m}_\alpha$, $\overline{\bm{w}}_\alpha$ and $\overline{r}_\alpha$. The simplest approach to specify these closures is through phenomenological arguments.  Alternatively, we can derive closure relations through homogenisation. In this regard, asymptotic homogenisation has been successful in deriving a closed dual-porosity model starting at the microscale (\citealt{Royer1994}). These results thus serve as a useful reference to the approach introduced herein. 

At the microscale we have the following mass balance
\begin{equation}
    \dot{m} + \nabla\cdot\bm{w} = 0, \label{eqn:7}
\end{equation}
where $m$ and $\bm{w}$ denote the rate of microscopic mass content change and mass flux vector respectively. Additionally, at this scale we assume both materials are saturated by the same slightly compressible fluid of density $\rho$. Under isothermal flows we define the fluid compressibility relation
\begin{equation}
    c_l = \frac{1}{\rho}\pderiv{\rho}{p}, \label{eqn:8}
\end{equation}
where $p$ denotes the microscopic pressure. A similar compressibility relation is defined for the local porosity $\phi=|\Omega^p|/|\Omega$|, where $\Omega^p$ is the pore volume, such that
\begin{equation}
    c_\phi = \frac{1}{\phi}\pderiv{\phi}{p}. \label{eqn:9}
\end{equation}
We assume small perturbations in fluid density and porosity allowing linearised evolutions of these quantities. Further, the small perturbation hypothesis permits the replacement of $\rho$ and $\phi$ with $\rho^0$ and $\phi^0$ respectively, where necessary. Notations with superscript $0$ denote reference conditions. From $m=\rho\phi$ and $\bm{w}=\rho\bm{q}$, where $\bm{q}$ is the microscopic volume flux, \cref{eqn:7} becomes 
\begin{equation}
    \rho^0\phi^0c\dot{p} + \nabla\cdot(\rho^0\bm{q}) = 0, \label{eqn:10}
\end{equation}
where we have made use of the small perturbation hypotheses and where $c=c_l+c_\phi$. Lastly, the microscopic volume flux is related to the microscopic pressure gradient through Darcy's law (neglecting gravitiational effects)
\begin{equation}
    \bm{q} = \frac{\mathsfbf{k}}{\mu}\cdot\nabla p, \label{eqn:11}
\end{equation}
where $\mu$ is the fluid viscosity. For the isotropic case $\mathsfbf{k}=\mathsf{k}\bm{I}$, in which $\bm{I}$ denotes the identity matrix. To link the macroscopic description in \crefrange{eqn:5}{eqn:6} to the microscale description in \cref{eqn:7} we make use of a two-scale homogenisation approach. This approach is split into two steps:
\begin{description}
    \item[Downscaling:] Involves translation of macroscopic loading parameters  (macroscopic pressures and gradients) to consistent constraints and boundary conditions at the microscale.
    \item[Upscaling:] Involves transformation of microscopic rates of change and fluxes to their macroscopic counterparts through averaging. 
\end{description}
We address the downscaling step in the section to follow. 

\subsection{Downscaling}
For a dual-porosity material, associated to a macroscopic point $\overline{\bm{x}}$ we have two macroscopic pressures $\overline{p}_1,\text{ }\overline{p}_2$ and gradients $\overline{\nabla}\overline{p}_1,\text{ }\overline{\nabla}\overline{p}_2$. For each constituent we then define microscopic pressure fields according to Taylor series expansions of the first order about the macroscopic point 
\begin{align}
    p(\overline{\bm{x}}, \bm{x}, t) &= \overline{p}_1(\overline{\bm{x}}, t) + \overline{\nabla}\overline{p}_1(\overline{\bm{x}}, t)\cdot(\bm{x}-\overline{\bm{x}}) + p'(\overline{\bm{x}}, \bm{x}, t) \quad \text{in } \Omega_1, \label{eqn:12} \\ 
    p(\overline{\bm{x}}, \bm{x}, t) &= \overline{p}_2(\overline{\bm{x}}, t) + \overline{\nabla}\overline{p}_2(\overline{\bm{x}}, t)\cdot(\bm{x}-\overline{\bm{x}}) + p'(\overline{\bm{x}}, \bm{x}, t) \quad \text{in } \Omega_2, \label{eqn:13}
\end{align}
where $p'$ denote higher order expansion terms corresponding to pressure fluctuations at the microscale. We note, the microscale is positioned relative to the macroscopic point such that $\av{\bm{x}-\overline{\bm{x}}}=0$, which also holds for constituent averages. 

From \crefrange{eqn:12}{eqn:13}, we require the following compatibilities to hold 
\begin{align}
    \av{p}_1 = \overline{p}_1, \label{eqn:14} \\
    \av{p}_2 = \overline{p}_2, \label{eqn:15}
\end{align}
corresponding to the requirement $\av{p'}_\alpha=0$. From, \cite{Ozdemir2008a}, \crefrange{eqn:14}{eqn:15} are related to requiring consistency between the microscopic and macroscopic mass contents such that 
\begin{equation}
    \overline{\rho}\overline{\phi}_\alpha\overline{c}_\alpha\overline{p}_\alpha = \av{\rho\phi cp_\alpha}, \label{eqn:16}
\end{equation}
where $\overline{\phi}_\alpha=|\Omega^p_\alpha|/|\Omega|=v_\alpha\phi_\alpha$ is the effective continuum porosity. From \cref{eqn:16}, we propose the following consistency conditions for the mass storage
\begin{equation}
    \overline{\rho}\overline{\phi}_\alpha\overline{c}_\alpha = \av{\rho\phi c}, \label{eqn:17}
\end{equation}
which is inline with results obtained through asymptotic homogenisation (\citealt{Auriault1983, Royer1994}). 

In addition to constraints provide by \crefrange{eqn:14}{eqn:15} we require averages of microscopic gradients to equal their macroscopic counterparts. Accordingly, taking the gradients of \crefrange{eqn:12}{eqn:13} and averaging over each constituent leads to 
\begin{align}
    \av{\nabla p}_1 &= \overline{\nabla}\overline{p}_1 + \av{\nabla p'}_1 \quad \text{in } \Omega_1, \label{eqn:18} \\ 
    \av{\nabla p}_2 &= \overline{\nabla}\overline{p}_2 + \av{\nabla p'}_2 \quad \text{in } \Omega_2, \label{eqn:19} 
\end{align}
where we use the following relation $\nabla(\bm{x}-\overline{\bm{x}})=\bm{I}$. To fulfill the compatibility between gradients we require $\av{\nabla p'}_\alpha=\bm{0}$ for both constituents. Application of Gauss's divergence theorem on averages of the fluctuation gradients leads to 
\begin{equation}
    \int_{\partial\Omega_\alpha}p'\bm{n} \text{ } dS=\bm{0}, \label{eqn:20}
\end{equation}
where $\partial\Omega_\alpha$ is the boundary of $\Omega_\alpha$ and $\bm{n}$ is the outward unit normal vector. Condition \cref{eqn:20} is satisfied by setting certain restrictions (boundary conditions) on the fluctuations at the boundaries. Examples of such boundary conditions include periodic boundary conditions and prescribed fluctuation boundary conditions. In this work we use the latter for demonstrative purposes, although the former is often used for flow-based upscaling procedures (\citealt{Lie2019}). Accordingly, prescribed fluctuation boundary conditions correspond to setting $p'=0$ leading to 
\begin{equation}
    p(\overline{\bm{x}}, \bm{x}, t) = \overline{p}_\alpha(\overline{\bm{x}}, t) + \overline{\nabla}\overline{p}_\alpha(\overline{\bm{x}}, t)\cdot(\bm{x}-\overline{\bm{x}}) \quad \text{on } \partial\Omega_\alpha, \label{eqn:21}
\end{equation}
from either \cref{eqn:12} or \cref{eqn:13}. In the dual-porosity formulation we assume the matrix material to be completely permeated by the fracture material. Additionally, we assume continuity of the microscopic pressure field at the interface $\Gamma=\partial\Omega_1$ between the two materials. Accordingly, boundary conditions for both materials can be completely defined by considering the prescribed fluctuation boundary conditions for the fracture material 
\begin{equation}
    p(\overline{\bm{x}}, \bm{x}, t) = \overline{p}_2(\overline{\bm{x}}, t) + \overline{\nabla}\overline{p}_2(\overline{\bm{x}}, t)\cdot(\bm{x}-\overline{\bm{x}}) \quad \text{on } \partial\Omega_2, \label{eqn:22}
\end{equation}
We note, \cref{eqn:22} is somewhat similar to the boundary conditions specified by the asymptotic homogenisation approach on $\Gamma$, where $p=\overline{p}_2 \text{ on } \Gamma$ (\citealt{Royer1994}). However, here the explicit size of the REV and macroscopic fracture pressure gradient is taken into account. Nonetheless, for small macroscopic gradients and vanishingly small REV sizes $p=\overline{p}_2 \text{ on } \Gamma$ is a reasonable approximation. With the required downscaling relations we can now proceed to upscaling. 

\subsection{Upscaling}
Upscaling for the macroscopic quantities ($\dot{\overline{m}}_\alpha, \overline{\bm{w}}_\alpha$) is achieved through a virtual power equivalence relation between the micro and macroscales. This relation is referred to as the generalised Hill-Mandel condition (\citealt{Blanco2016}), and requires the macroscopic virtual power to equal its volume averaged microscopic counterpart at a given material point. Here we introduce a virtual power relation for the dual-porosity material by summing the external powers coming from \crefrange{eqn:5}{eqn:6} and using $\delta p_1\overline{r}_1+\delta p_2\overline{r}_2=0$ to give
\begin{equation}
    \delta\overline{p}_1\dot{\overline{m}}_1-\overline{\nabla}\delta\overline{p}_1\cdot{\overline{\bm{w}}}_1 + \delta\overline{p}_2\dot{\overline{m}}_2-\overline{\nabla}\delta\overline{p}_2\cdot{\overline{\bm{w}}}_2 = \av{\delta p\dot{m}-\nabla\delta p\cdot\bm{w}}, \label{eqn:23}
\end{equation}
where we use \cref{eqn:7} at the microscale for compactness. Use of the decomposition in \cref{eqn:3} on \cref{eqn:23} followed by substitution of the appropriate expansions in \crefrange{eqn:12}{eqn:13} leads to 
\begin{align}
    \sum_{\alpha=1,2}\delta\overline{p}_\alpha\dot{\overline{m}}_\alpha-\overline{\nabla}\delta\overline{p}_\alpha\cdot{\overline{\bm{w}}}_\alpha &= v_1\langle[\delta\overline p_1 + \overline{\nabla}\delta\overline{p}_1\cdot(\bm{x}-\overline{\bm{x}})+ \delta p']\dot{m} \nonumber \\
    &-[\nabla\delta \overline{p}_1+\nabla\delta p']\cdot\bm{w} \rangle_1 \nonumber \\
    &+v_2\langle[\delta\overline p_2 + \overline{\nabla}\delta\overline{p}_2\cdot(\bm{x}-\overline{\bm{x}})+ \delta p']\dot{m} \nonumber \\
    &-[\nabla\delta \overline{p}_2+\nabla\delta p']\cdot\bm{w}\rangle_2. \label{eqn:24}
\end{align}
Rearranging \cref{eqn:24} to isolate terms involving fluctuations gives 
\begin{align}
    \sum_{\alpha=1,2}\delta\overline{p}_\alpha\dot{\overline{m}}_\alpha-\overline{\nabla}\delta\overline{p}_\alpha\cdot{\overline{\bm{w}}}_\alpha &= v_1\langle\delta\overline{p}_1\dot{m}-\overline{\nabla}\delta\overline{p}_1\cdot[{\bm{w}+(\bm{x-\overline{\bm{x}}})\dot{m}]\rangle}_1 \nonumber \\
    &+ v_2\langle\delta\overline{p}_2\dot{m}-\overline{\nabla}\delta\overline{p}_2\cdot[{\bm{w}+(\bm{x-\overline{\bm{x}}})\dot{m}]\rangle}_2 \nonumber \\
    &+ v_1\langle\delta p' \dot{m}-\nabla\delta p'\cdot\bm{w}\rangle_1 \nonumber \\
    &+ v_2\langle\delta p' \dot{m}-\nabla\delta p'\cdot\bm{w}\rangle_2. \label{eqn:25}
\end{align}
The terms in \cref{eqn:25} involving the pressure fluctuations represents the weak form of the phase-wise microfluctuation balance of mass. These terms are subsequently addressed through the chain rule and the divergence theorem leading to 
\begin{align}
    v_1\langle\delta p'\dot{m}&-\nabla\delta p'\cdot\bm{w}\rangle_1 + v_2\langle\delta p'\dot{m}-\nabla\delta p'\cdot\bm{w}\rangle_2 \nonumber \\
    &=\langle \delta p'(\dot{m}+\nabla\cdot\bm{w})\rangle - \int_{\partial\Omega_1}p'\bm{n} \text{ } dS - \int_{\partial\Omega_2}p'\bm{n} \text{ } dS. \label{eqn:26}
\end{align}
Accordingly, the first term on the right-hand side of \cref{eqn:26} is satisfied due to the microscopic conservation of mass \cref{eqn:7}. The remaining terms then vanish with the appropriate boundary conditions, in this case the prescribed fluctuation boundary conditions. 

Having dealt with the fluctuation terms we seek to obtain a dual-porosity formulation from the remaining terms in \cref{eqn:25}. This goal is pursued in the following section. 

\subsection{Recovery of the dual-porosity model}
To recover a dual-porosity model from the virtual power equivalence we assume quasi-steady state in the fracture material due to the high phase permeability. Accordingly,
\begin{equation}
    \nabla\cdot(\bm{w}) = \rho^0\nabla\cdot(\bm{q})=0 \quad \text{in } \Omega_2, \label{eqn:27}
\end{equation}
whilst transient diffusion effects remain in the matrix material at the microscale. However, due to the percolating fractures and the low intrinsic matrix permeability we assume $\av{w}_1\approx0$. With the following assumptions at the microscale we recover from \cref{eqn:25} 
\begin{align}
    \delta\overline{p}_1\dot{\overline{m}}_1-\overline{\nabla}\delta\overline{p}_1\cdot{\overline{\bm{w}}}_1+\delta\overline{p}_2\dot{\overline{m}}_2-\overline{\nabla}\delta\overline{p}_2\cdot{\overline{\bm{w}}}_2
    &= \delta\overline{p}_1\langle{\dot{m}\rangle}_1v_1 \nonumber \\
    &-\overline{\nabla}\delta\overline{p}_1\cdot\langle{(\bm{x-\overline{\bm{x}}})\dot{m}\rangle}_1v_1 \nonumber \\
    &-\overline{\nabla}\delta\overline{p}_2\cdot\langle{\bm{w}\rangle}_2v_2. \label{eqn:28}
\end{align}
Comparing terms in \cref{eqn:28} we identify the following relations 
\begin{align}
    \dot{\overline{m}}_1 &= v_1\langle{\dot{m}\rangle}_1=\rho^0\overline{\phi}^0_1c_1\av{\dot{p}}_1, \label{eqn:29} \\
    \overline{\bm{w}}_1 &= v_1\av{(\bm{x-\overline{\bm{x}}})\dot{m}}_1, \label{eqn:30} \\
    \overline{\bm{w}}_2 &= v_2\av{\bm{w}}_2 = \rho^0v_2\av{\bm{q}}_2, \label{eqn:31}
\end{align}
where fluid density at reference conditions at the microscale corresponds to the equivalent quantity at the macroscale i.e. $\overline{\rho}^0=\rho^0$. Further, for the small perturbation considerations used here $\overline{c}_\alpha = c_\alpha$.

Looking first at the averaged quantity in \cref{eqn:30}, we identify this as the first moment of the rate of pressure change in the matrix. Interestingly, this term is not present in the classical asymptotic homogenisation approach (\citealt{Royer1994}). Physically, this term introduces a size-effect and represents the microscale mass inertia due to the low permeability of the matrix material. However, in general this inertial term is shown to be negligible with respect to the macroscopic flux from the fractures (\citealt{Brassart2019}). Further, for isotropically distributed materials at the microscale this term is shown to vanish (\citealt{Brassart2019}). 

From \cref{eqn:31} we can identify the macroscopic volumetric fracture flux as 
\begin{equation}
    \overline{\bm{q}}_2 = v_2\av{\bm{q}}_2\approx\av{\bm{q}}, \label{eqn:32}
\end{equation}
where the relation on the right-hand side of \cref{eqn:32} arises due to the assumption $v_1\av{\bm{q}}_1\approx0$ and from \cref{eqn:3}. Next, Using \cref{eqn:11} in \cref{eqn:32} we write
\begin{equation}
    \overline{\bm{q}}_2 = \frac{\overline{\mathsfbf{k}}_2}{\mu}\cdot\overline{\nabla}\overline{p}_2, \label{eqn:33}
\end{equation}
where $\overline{\mathsfbf{k}}_2$ denotes the (macroscopic) effective fracture permeability tensor. Accordingly, we can recover entries to this permeability tensor provided we can calculate the macroscopic fracture flux corresponding to the macroscopic pressure gradient. Such calculations are addressed in so-called flow-based upscaling procedures (\citealt{Lie2019}).

Finally, \cref{eqn:29} shows the relation between the micro and macroscopic rate of mass content change. From \cref{eqn:4} we see 
\begin{equation}
    \rho^0\overline{\phi}^0_1c_1\av{\dot{p}}_1 = \rho^0\overline{\phi}^0_1c_1\dot{\av{p}}_1. \label{eqn:34}
\end{equation}
Accordingly, with \cref{eqn:29} (resp. \cref{eqn:34}) and \cref{eqn:31} in \crefrange{eqn:5}{eqn:6}, whilst neglecting the inertial effects from \cref{eqn:30}, we recover the dual-porosity model from the two-scale homogenisation procedure as
\begin{align}
        \rho^0\overline{\phi}^0_1c_1\dot{\av{p}}_1 &= \overline{r}_1, \label{eqn:35} \\
        \rho^0\overline{\phi}^0_2c_2\dot{\av{p}}_2 + \rho^0\overline{\nabla}\cdot(v_2\av{\bm{q}}_2) &= -\overline{r}_1, \label{eqn:36}
\end{align}
where we use the compatibility in \cref{eqn:16} for the fracture rate of mass content change $\dot{\overline{m}}_2$. 

In this work our focus is in the modelling of the inter-porosity transfer term $\overline{r}_1$. From \cref{eqn:35} we see $\overline{r}_1=\rho^0\overline{\phi}^0_1c_1\dot{\av{p}}_1$. Subsequently, to compute $\overline{r}_1$ \textit{all} we need is the solution to the microscale boundary value problem \cref{eqn:10} in the matrix, given boundary conditions from \cref{eqn:22} on $\Gamma$ and some initial condition $p_1(\overline{\bm{x}}, \bm{x}, t=0)=p^0_1$. However,
often we not have direct access to the microscale within a macroscopic simulation. Further, even if the solution could be calculated directly a-priori, it is difficult to use in practice owing to its complex form. Specifically, for evolving fracture pressures, the matrix pressure solution is defined as a convolution product over the fracture pressure history (\citealt{Royer1994}). Even for static fracture pressures following a step increase, the matrix pressure solution is still an infinite series defined in explicit time (shown in the following section). Due to the described complexities, it is common to adopt a simple linear relation to model $\overline{r}_1$. Such simplification leads to measurable errors in flow behaviour. Accordingly, in the remainder of this work we aim to improve on linear transfer models through machine learnt constitutive models which incorporate temporal information using quantities directly available during simulation.  

\section{A dual-porosity model problem and recovery of a first-order inter-porosity mass transfer model}
With the homogenisation step in-hand we now want to proceed to the remaining steps of the multiscale framework. To do so, we first consider a model problem for which data can be generated using analytical solutions. Further, from these solutions we can obtain a first-order mass transfer model that is commonly used in dual-continuum flow modelling. This simplified transfer model will then serve as a reference for testing any data-driven model coupled to a numerical simulator. We note that whilst the problem considered here is not too complex, it serves as a useful demonstration for the remaining steps of the framework and to explore various established machine learning algorithms for modelling the time-dependent mass transfer problem. Future directions therefore concern numerically derived data involving fracture dynamics, more complicated microscale geometries, and more advanced methods for surrogate modelling (e.g. active learning, physics-informed models etc.). In this latter element, work combining the data-driven parameterisations from \cite{Andrianov2021} with elements of the framework introduced herein could be an interesting path for further research.  

\subsection{Model problem}
\begin{figure}[h]
\centering
\begin{minipage}[b]{0.5\textwidth}
\centering
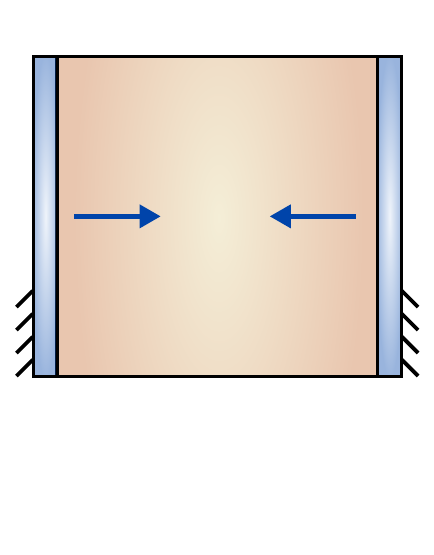
\subcaption{Problem geometry} \label{fig:2a}
\end{minipage}%
\begin{minipage}[b]{0.5\textwidth}
\centering
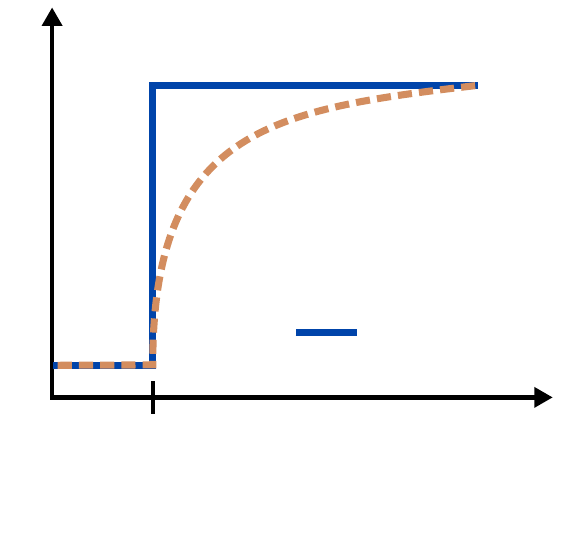
\subcaption{Pressure step} \label{fig:2b}
\end{minipage}%
\caption{Model problem setup: Geometry (a), and fracture pressure step (b).}
\label{fig:2}
\end{figure}

For the model problem we consider a two-dimensional matrix domain surrounded by fractures (\cref{fig:2a}). At a time $t^0$ the matrix and fractures are in hydrostatic equilibrium. Then, following a jump in fracture pressure at a time $t^{0+}$, the fracture pressure remains fixed leading to $\overline{\nabla}\overline{p}_2=0$ (\cref{fig:2b}). Accordingly, from \cref{eqn:21} and the pressure continuity across the interface we have 
\begin{equation}
    p(\overline{\bm{x}}, \bm{x}, t) = \overline{p}_2(\overline{\bm{x}}, t) \quad \text{on } \partial\Omega_1. \label{eqn:37}
\end{equation}
with initial conditions
\begin{equation}
    p(\overline{\bm{x}},\bm{x},t=0) = p^0_1(\overline{\bm{x}}, \bm{x}) \quad \text{in } \Omega_1. \label{eqn:38}
\end{equation}
Due to the considered (fracture pressure) boundary conditions and simple geometry, analytical solutions exist for calculating the solution. These solutions are introduced subsequently. 

\subsection{Analytical solution and first-order mass transfer model}
The analytical solution to the 2D matrix diffusion problem described above is found by taking the product of the (dimensionless) solution to a 1D diffusion problem such as those arising in slab geometries (\citealt{Crank1979, Holman1990, Lim1995}). Averaging the solution over the REV, and noting the equivalence with macroscopic quantities, then leads to
\begin{align}
    \overline{p}_1 = p^0_1 + (\overline{p}_2-p^0_1)\Bigg\{&1-\sum_{n=0}^{\infty}\sum_{m=0}^{\infty}\left(\frac{8}{\pi^2}\right)^2\frac{1}{(2n+1)^2(2m+1)^2} \nonumber \\
    &\text{exp}\left(-\frac{\pi^2\mathsf{k}_1t}{\phi^0_1c_1 \mu\ell^2}\left[(2n+1)^2+(2m+1)^2\right]\right)\Bigg\}, \label{eqn:39}
\end{align}
where we use $\overline{p}_1=\av{p}_1$. Next, taking the time derivative of \cref{eqn:39} gives
\begin{align}
    \dot{\overline{p}}_1 =  (\overline{p}_2-p^0_1)\sum_{n=0}^{\infty}\sum_{m=0}^{\infty}&\left( \frac{8}{\pi^2}\right)^2\frac{\pi^2\mathsf{k}_1}{\phi^0_1c_1\mu \ell^2}\left[\frac{1}{(2n+1)^2}+\frac{1}{(2m+1)^2}\right] \nonumber \\
    &\text{exp}\left(-\frac{\pi^2\mathsf{k}_1t}{\phi^0_1c_1\mu \ell^2}\left[(2n+1)^2+(2m+1)^2\right]\right). \label{eqn:40}
\end{align}
Substitution of \cref{eqn:40} into \cref{eqn:35} gives $\overline{r}_1$. However, the resulting mass transfer expression is unsuitable for simulation purposes given the presence of explicit time and the infinite series. To alleviate these dependencies we can recover a first-order approximation to \cref{eqn:40}. Accordingly, taking the first term from \cref{eqn:40} and eliminating $t$ using the first term from \cref{eqn:39} gives
\begin{equation}
    \dot{\overline{p}}_1 = \frac{2\pi^2\mathsf{k}_1}{\phi^0_1 c_1\mu\ell^2}(\overline{p}_2-\overline{p}_1). \label{eqn:41}
\end{equation}
Substitution of \cref{eqn:41} into \cref{eqn:35} leads to the following first-order inter-porosity transfer model (\citealt{Zimmerman1993, Lim1995})
\begin{equation}
    \overline{r}_1 = \rho^0\frac{v_1\eta \mathsf{k}_1}{\mu}(\overline{p}_2-\overline{p}_1), \label{eqn:42}
\end{equation}
where $\eta=(N_f\pi^2)/\ell^2$ denotes the so-called shape factor in which $N_f$ is the number of characteristic fracture sets. 

Several investigations have shown use of the linear constitutive model in \cref{eqn:42} to lead to measurable errors in flow behaviour (\citealt{Zimmerman1993, Ashworth2020}). One common way to alleviate these inaccuracies is to use multi-rate transfer models (\citealt{Haggerty1995, Geiger2013}). However, in this work we treat the problem differently. Instead, we use machine learning to bridge the scales in the form of a surrogate constitutive model. 

\section{Application of the framework: Data generation}
Our goal for the supervised learning problem is to learn a mapping from quantities available during simulation to an output aligned with our constitutive property. To do so we consider a dataset $\mathcal{D}$ of $n$ samples, $\mathcal{D}=\{(\mathsfbf{x}_i,\mathsfbf{y}_i)\text{ }|\text{ }i=1,...,n\}$, where $\mathsfbf{x}_i\in\mathbb{R}^{d_{in}}$ is a vector of inputs and $\mathsfbf{y}_i\in\mathbb{R}^{d_{out}}$ is a (vector) quantity derived from the microscale. Learning a mapping between inputs and outputs then equates to learning a function of the form
\begin{equation}
    \mathsfbf{y}_i = f(\mathsfbf{x}_i) \approx \hat{f}(\mathsfbf{x}_i) \label{eqn:43}.
\end{equation}
As a result, we can incorporate local scale physics, such as those coming from the analytical solution, or numerical simulations, within macroscopic models without having to formulate explicit phenomenological expressions.

\begin{figure}[h]
\centering
\begin{minipage}[b]{0.5\textwidth}
\centering
    \begin{tikzpicture}
        \begin{loglogaxis}[
            xminorticks=true,
            xmin=0.1, xmax=10e1,
            yminorticks=true,
            ymin=10, ymax=10e5,
            xlabel = Time (s),
            xlabel style={font=\fontsize{8}{144}\selectfont\color{white!15!black}},
            ylabel = $\overline{p}_1$ (Pa),
            ylabel style={font=\fontsize{8}{144}\selectfont\color{white!15!black}},
            ticklabel style={font=\fontsize{8}{144}},
            width=4.5cm, height=4.5cm,
            scale only axis,
            enlargelimits=false,
            axis on top]
            \addplot graphics[xmin=0.1, xmax=10e1, ymin=10, ymax=10e5] {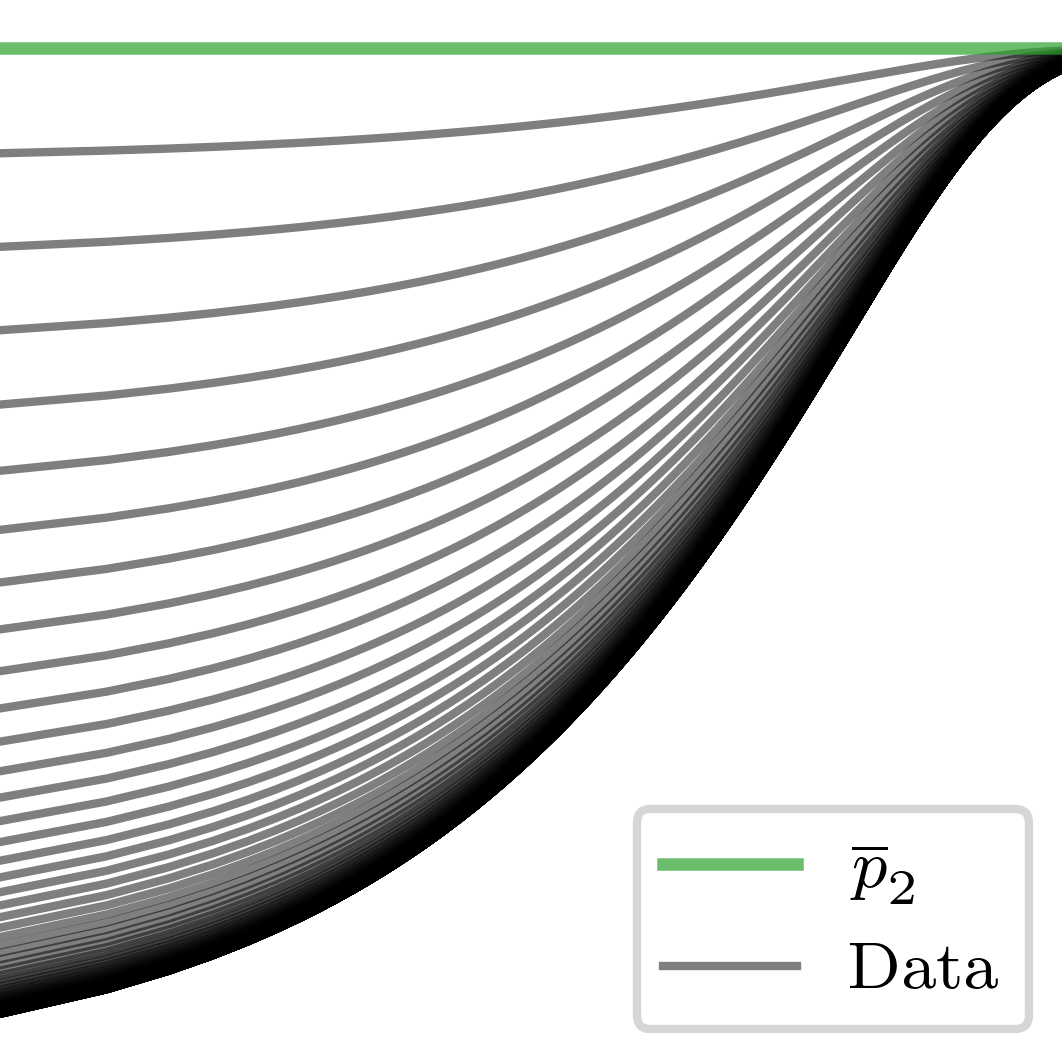};
        \end{loglogaxis}
  \end{tikzpicture}
  \subcaption{Time series data}
  \label{fig:3a}
\end{minipage}%
\begin{minipage}[b]{0.5\textwidth}
\centering
    \begin{tikzpicture}
        \begin{loglogaxis}[
            xminorticks=true,
            xmin=0.1, xmax=10e1,
            yminorticks=true,
            ymin=10, ymax=10e5,
            xlabel = Time (s),
            xlabel style={font=\fontsize{8}{144}\selectfont\color{white!15!black}},
            ylabel = $\overline{p}_1$ (Pa),
            ylabel style={font=\fontsize{8}{144}\selectfont\color{white!15!black}},
            ticklabel style={font=\fontsize{8}{144}},
            width=4.5cm, height=4.5cm,
            scale only axis,
            enlargelimits=false,
            axis on top]
            \addplot graphics[xmin=0.1, xmax=10e1, ymin=10, ymax=10e5] {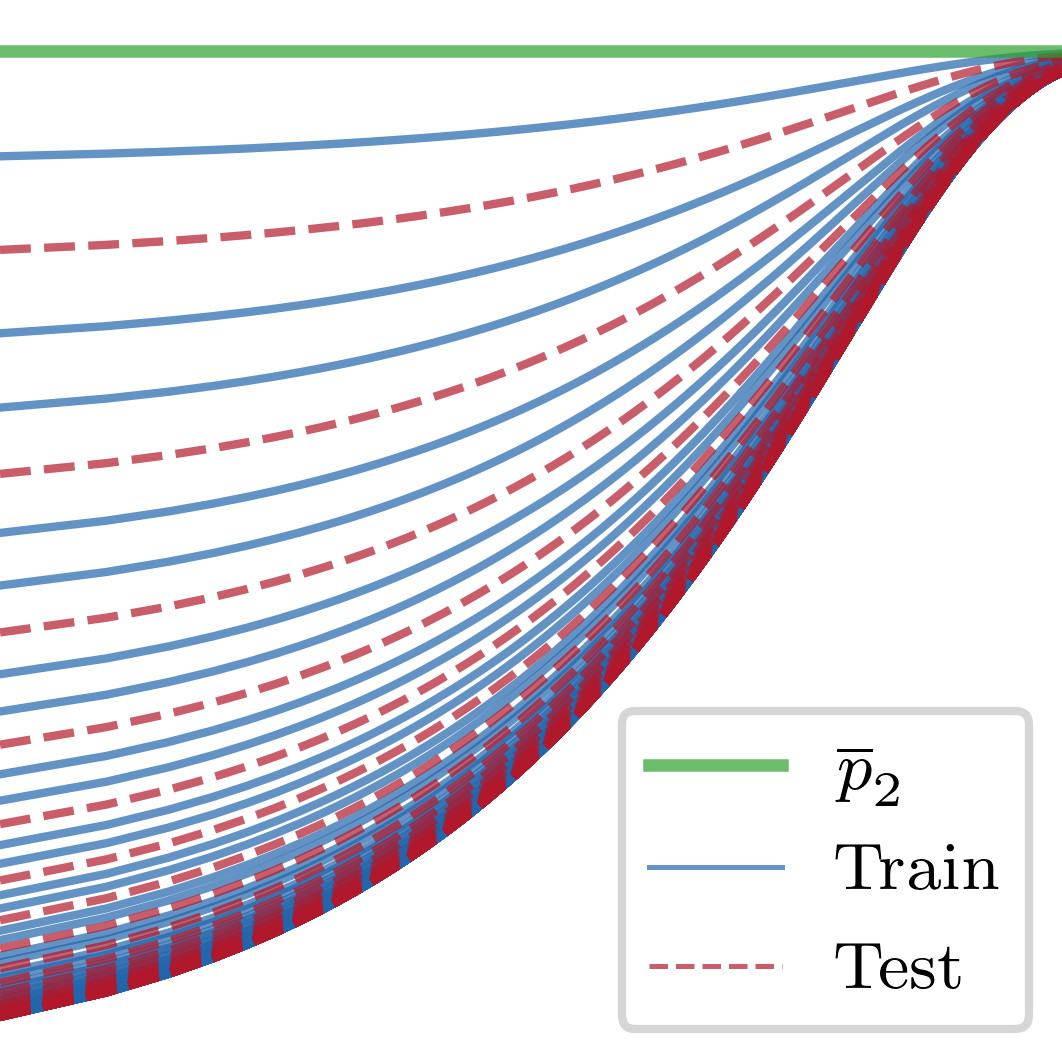};
        \end{loglogaxis}
  \end{tikzpicture}
  \subcaption{Train and test data split} \label{fig:3b}
\end{minipage}
\caption{Time series data for the learning problem: Raw data (a), and training and test data split (b).} \label{fig:3}
\end{figure}

We frame our learning problem considering the physics of the process, inputs coming from simulation and an output that is aligned with the discrete structure of the numerical problem. For the latter, we specify our output following application of the backward Euler method commonly used for time discretisation in numerical modelling. Specifically 
\begin{equation}
    \pderiv{\overline{p}_1}{t} \approx  \frac{\overline{p}_1(t+1)-\overline{p}_1(t)}{\Delta t}. \label{eqn:44}
\end{equation}
From \cref{eqn:44}, our goal is to predict the pressure at the next time step
\begin{equation}
    \mathsf{y}_i(t+1) = \overline{p}_1(t+1). \label{eqn:45}
\end{equation}
At simulation time we incorporate the resulting prediction into the finite-difference calculation in \cref{eqn:44} at each time step. Accordingly, our learning problem is formulated on the basis of time series calculated with \cref{eqn:39}. Each individual time series corresponds to a different initial matrix pressure $p^0_1$. We take 120 different $p^0_1$, sampled with logarithmic spacing from the interval of $[1 , 1\times10^6] \text{ Pa}$. Every series starts from $0$ and finishes at a final time of $T=100 \text{ s}$, with a uniform stepping interval of $0.1\text{ s}$. The remaining parameters in \cref{eqn:39} are fixed as $\mathsf{k}_1 = 0.5 \text{ md}$, $\phi^0_1 = 0.2$, $c_1=1.4 \text{ GPa}^{-1}$, $\mu = 1 \text{ cp}$, $\ell = 1 \text{ m}$, and following the step in $\overline{p}_2$ at $t^{0+}$, $\overline{p}_2 = 1 \text{ MPa}$. A plot of each series $s$ as time vs $\mathsfbf{y}^s$ is shown in \cref{fig:3a}. Further, \cref{fig:3b} shows how this time series data is split into training and testing data using a $2/3:1/3$ split respectively. Accordingly, every third sequence corresponds to a test series. 

\section{Application of the framework: Surrogate constitutive model learning}
In this section we review and compare the learning algorithms used to generate our surrogate constitutive model. Specifically, we introduce the various approaches and considerations for modelling the time-dependent problem. To follow, we train and test the different modelling approaches on our data. The results from this section will then be carried on to the next stage of the framework i.e. coupling the data-driven model to a numerical simulator. 

\subsection{Autoregressive approaches}
To start, we introduce the modelling description used to account for time-dependency. That is, we model the sequential problem using an autoregressive description. In the following, we present autoregressive descriptions in the context of feedforward models. In feedforward approaches the information flow is unidirectional from the current inputs to outputs. This modelling approach is contrary to recurrent models, in which the information flow comes from both the current inputs and from calculations made on inputs at previous time steps. 

For a (multivariable) feedforward autoregressive model, $\hat{f}(\cdot)$ uses previous series outputs to predict the next output in the series. Accordingly, for a two variable, univariate feedforward autoregressive model
\begin{align}
    \hat{\mathsf{y}}_i(t+1) = \hat{f}\left(\mathsfbf{x}_i(t)\right) = \hat{f}([&\hat{\mathsf{y}}_i(t),...,\hat{{\mathsf{y}}}_i(t-d_{\mathsf{y}}+1), \nonumber \\ &\mathsf{u}_i(t),...,\mathsf{u}_i(t-d_\mathsf{u}+1)]), \label{eqn:46}
\end{align}
where $\mathsf{u_i}$ denotes an external series, and $d_\mathsf{y}$ and $d_\mathsf{u}$ denote the number of sequence terms, or dependency length, for the internal and external series variables respectively. In this work we set $d_\mathsf{y}=d_\mathsf{u}=d$. 

Our external series corresponds to the boundary conditions for the matrix provided by the homogenisation approach given the model problem described in Section 4. Specifically, previous sequence values for $\overline{p}_2$ corresponding to $\overline{p}_2=p^0_1$ when $t<t^{0+}$, and $\overline{p}_2=1 \text{ MPa}$ (or any selected pressure step value) when $t\geq t^{0+}$. Finally, we note that $d_\mathsf{y}$ and $d_\mathsf{u}$ (or $d$) are defined a-priori. As a result, potentially important long-range information is lost if short dependency lengths are chosen.  To alleviate the challenges posed by fixed dependency lengths one could use recurrent neural networks such as the long short-term memory network. However, for the purposes of this work, we save such experiments for future studies.  

For feedforward autoregressive training we treat the learning task as a \textit{single-step ahead} prediction problem. In this strategy, the idea is to use lagged ground truths (targets) $\mathsf{y}_i$ in place of the predicted values $\hat{\mathsf{y}}_i$ in \cref{eqn:46}. 

Autoregressive testing is treated according to how the ML model will be used when coupled to the (macroscopic) simulator. That is, as a \textit{multi-step ahead} prediction problem. Accordingly, the trained single-step ahead models are used in a recursive manner, with predictions $\hat{\mathsf{y}}$ at previous time steps fed back into the inputs to predict the next time step, as per \cref{eqn:46}. The single-step ahead training and testing strategy can work provided the training error is sufficiently small such that accumulated errors do not explode when using the model as a multi-step ahead predictor. When this requirement on the accumulated errors cannot be satisfied (which is difficult to determine a-priori), iterative approaches are unstable. In such cases, multi-output strategies that predict the whole forecast horizon at once can be used. However, the stability of multi-output regressors often comes at the cost of predictive accuracy, particularly for long forecast horizons (\citealt{Sangiorgio2020}).

In the section to follow we consider two ML algorithms for feedforward autoregressive modelling. Specifically, polynomial regression (PR) and a fully connected neural network (FC-NN). 

\subsection{Polynomial regression}
Here we introduce the polynomial regression algorithm. These aglorithms are attractive because they are easy to understand and fast to train. However, a noteable downside to these methods is the curse of dimensionality for large input vectors. Nonetheless, PR algorithms are a simple place to start for many regression tasks, and serve as a good baseline model. Accordingly, under PR $\hat{f}(\cdot)$ is formulated as
\begin{equation}
    \hat{\mathsf{y}}_i = \hat{f}(\mathsfbf{x}_i) = \mathsfbf{w}^\top\bm{\theta}({\bf x}_i), \label{eqn:47}
\end{equation}
where $\bm{\theta}=[\theta_1,...,\theta_{n_p}]$ is the vector of polynomial basis functions acting on $\mathsfbf{x}_i$ and $\mathsfbf{w}\in\mathbb{R}^{n_p}$ is a vector of weights (parameters) of the model. In this work we use a second order polynomial (feature) transformation, which comes as a standard tool in many machine/statistical learning packages. 

Whilst the inputs are polynomial (following a transformation) in \cref{eqn:47}, the weights remain linear. Finding $\mathsfbf{w}$ is simply then a linear regression problem and is the goal of training. In supervised learning problems, training corresponds to minimising a loss function $\mathcal{L}(\cdot)$ that represents the error in approximating $f(\cdot)$ by $\hat{f}(\cdot)$ on our data. For the regression problem we use a squared loss such that 
\begin{equation}
     \mathcal{L}(\mathsfbf{w}) = \frac{1}{n}\|\mathsfbf{y}-\hat{f}(\bm{\Theta};\mathsfbf{w})\|_2^2, \label{eqn:48}
\end{equation}
where $\bm{\Theta}=[\bm{\theta}({\bf x}_1), \bm{\theta}(\mathsfbf{x}_2), ..., \bm{\theta}(\mathsfbf{x}_n)]^\top$ and $\mathsfbf{y}=[\mathsf{y}_1,\mathsf{y}_2,...,\mathsf{y}_n]^\top$ for the univariate target case.
Finding the optimal set of weights that minimise \cref{eqn:48} is defined formally as
\begin{equation}
    \argmin_{\mathsfbf{w}}\mathcal{L}(\mathsfbf{w}) = \argmin_{\mathsfbf{w}}\frac{1}{n}\|\mathsfbf{y}-\hat{f}(\bm{\Theta};\mathsfbf{w})\|_2^2. \label{eqn:49}
\end{equation}
The squared loss in linear regression is a convex function which admits a global minimum in $\mathsfbf{w}$ at the point $\nabla\mathcal{L}=0$. Accordingly
\begin{equation}
    \bm{\Theta}^\top\bm{\Theta}\mathsfbf{w} = \bm{\Theta}^\top\mathsfbf{y}, \label{eqn:50}
\end{equation}
leading to
\begin{equation}
    \mathsfbf{w} = \bm{\Theta}^\dagger\mathsfbf{y}, \label{eqn:51}
\end{equation}
where $\bm{\Theta}^\dagger=(\bm{\Theta}^\top\bm{\Theta})^{-1}\bm{\Theta}^\top$.

PR is well known to suffer from high degrees of multicollinearity (\citealt{Dalal2012}). As a result, $\bm{\Theta}^\top\bm{\Theta}$ is ill-conditioned and the solution in \cref{eqn:52} is non-unique. For prediction problems such ill-conditioning is generally not a concern (\citealt{Dormann2013}), since new data comes from the same population as the training data.  However, for problems of extrapolation such as multi-step ahead prediction, collinearity between inputs can change due to accumulating errors. Such changes can affect predictions on models arising from ill-conditioned matrices and should be addressed (\citealt{Dormann2013}). We therefore address multicollinearity between inputs using regularisation, which aims to restrict the space of viable models that fit our data (e.g. in the face of ill-conditioning). There are various forms of regularisation, however in this work we add the following penalisation $\gamma\|\mathsfbf{w}\|^2_2$ term to \cref{eqn:48} leading to so-called ridge regression. The parameter $\gamma$ in the penalty then controls the degree of weight penalisation. 

\subsection{Fully connected neural networks}
Here we introduce the fully connected neural network model. Aside from the FC-NN considered in this work, neural networks more generally have become a popular choice for surrogate modelling due to their great flexibility, scalability and significant advancement within research communities. Further, contrary to the polynomial regression where the feature transforms are defined a-priori, in neural networks these transformations are implicit within the algorithm and learnt from the data. 

To proceed, within an FC-NN information is fed forward from an input layer through hidden layers ($l=1,...,L-1$) finishing at an output layer ($l=L$). Accordingly, under an FC-NN $\hat{f}(\cdot)$ is formulated as
\begin{equation}
    \hat{\mathsf{y}}_i = \hat{f}(\mathsfbf{x}_i) = \mathsf{s}^{[L]}\left(\mathsf{s}^{[L-1]}\left(...\left(\mathsf{s}^{[1]}\left(\mathsfbf{x}_i\right)\right)...\right)\right). \label{eqn:52}
\end{equation}
For a given layer $l$, $\mathsf{s}^{[l]}$ is given as an affine transformation followed by an element-wise nonlinear function such that
\begin{equation}
    \mathsf{s}^{[l]}(\mathsfbf{x}) = h^{[l]}\left(\mathsfbf{W}^{[l]\top}\mathsfbf{x}+\mathsfbf{b}^{[l]}\right), \label{eqn:53}
\end{equation}
where $\mathsfbf{W}^{[l]}\in\mathbb{R}^{n_{l-1}\times n_{l}}$ and $\mathsfbf{b}^{[l]}\in\mathbb{R}^{n_l}$ are the weight matrix and bias parameter vector associated with layer $l$ respectively. Evidently, for the univariate prediction considered herein, the weight matrix for the last layer will be a weight vector $\mathsfbf{W}^{[L]}=\mathsfbf{w}^{[L]}$. Notation $h(\cdot)$ is the nonlinear function referred to as an activation function. In this work, we take $h^{[L]}$ as a linear activation and all others as so-called exponential linear units. In training the FC-NN we minimise the customary squared loss of the form
\begin{equation}
     \mathcal{L}(\bm{\lambda}) = \frac{1}{n}\|\mathsfbf{y}-\hat{f}(\mathsfbf{X};\bm{\lambda})\|_2^2, \label{eqn:54}
\end{equation}
where $\bm{\lambda}=[\mathsfbf{W}^{[1]},\mathsfbf{b}^{[1]},...,\mathsfbf{W}^{[L]},\mathsfbf{b}^{[L]}]$ and $\mathsfbf{X}=[\mathsfbf{x}_1,\mathsfbf{x}_2,...,\mathsfbf{x}_n]^\top$. However, unlike \cref{eqn:48}, with neural networks the loss function is non-convex. As a result, weights and biases are updated iteratively using a chosen optimiser together with the backpropogation algorithm. In this work, we found stochastic gradient descent with Nesterov momentum to work the best. We experimented with various model depths and widths, and found the best results came using a simple single hidden layer FC-NN with 12 hidden units. Additionally, with this architecture there was little overfitting for the single-step problem, and hence no need for regularisation. Our FC-NN was implemented using the Keras API (\citealt{Chollet2015}). 

Lastly, we construct the datasets used by these models as $\mathcal{D}=(\mathsfbf{X}, \mathsfbf{y}$ such that $\mathsfbf{X}=[\mathsfbf{x}^1_1,\mathsfbf{x}^1_{2},...,\mathsfbf{x}^S_{1000}]^\top$ and $\mathsfbf{y}=[\mathsf{y}^1_1,\mathsf{y}^1_{2},...,\mathsf{y}^S_{1000}]^\top$. Superscripts on the input matrix and target vector entries denote the time series label, with $S(=120)$ being the total number of series. Subscripts then denote the time step within a given series. 

\subsection{Training and testing}
In the following we present some additional considerations for model training and interpretation. Then, we present the results and discussions for the model training (single-step ahead) and testing (multi-step ahead) procedures.

\subsubsection*{Considerations}
For our models we consider a dependency length of $d=3$ for the inputs. Although not studied in detail here, $d$ is a hyperparameter that can be tuned according to the trade-off between gain in accuracy and complexity. 

The weight penalty term $\gamma$ for the regularised polynomial regression was found to give the best results for $\gamma=1\times10^{-7}$. We present results comparing the impact of different values for $\gamma$ for the multi-step ahead problem on both training and testing sets in the section to follow. 

When training our FC-NN model we used a batch size of 128, run over 500 epochs. Further, we standardise our data to have a mean of zero and a standard deviation of one. Standardisation during testing is then done using the mean and standard deviation calculated on the training set. 

To evaluate between the different models we consider one minus the normalised root-mean-square error given as
\begin{equation}
    1-NRMSE = 1-\frac{RMSE}{\sigma}= \frac{\sqrt{(n)^{-1}\sum_{i=1}^n(\mathsf{y}_i-\overline{\mathsf{y}})^2}}{\sqrt{(n-1)^{-1}\sum_{i=1}^n(\mathsf{y}_i-\overline{\mathsf{y}})^2}}, \label{eqn:55}
\end{equation}
where $\sigma$ and $\overline{\mathsf{y}}$ represent the sample standard deviation and mean of the (train or test) target set respectively. From \cref{eqn:55} the closer the value to one, the more accurate the algorithm. Lastly, to highlight the effects of multicollinearity, we present results for the polynomial regression model both with and without regularisation.

\subsubsection*{Results}
\begin{figure}[h]
\centering
\begin{minipage}[b]{0.5\textwidth}
\centering
    \begin{tikzpicture}
        \begin{loglogaxis}[
            xminorticks=true,
            xmin=0.1, xmax=10e1,
            yminorticks=true,
            ymin=10, ymax=10e5,
            xlabel = Time (s),
            xlabel style={font=\fontsize{8}{144}\selectfont\color{white!15!black}},
            ylabel = $\overline{p}_1$ (Pa),
            ylabel style={font=\fontsize{8}{144}\selectfont\color{white!15!black}},
            ticklabel style={font=\fontsize{8}{144}},
            width=4.5cm, height=4.5cm,
            scale only axis,
            enlargelimits=false,
            axis on top]
            \addplot graphics[xmin=0.1, xmax=10e1, ymin=10, ymax=10e5] {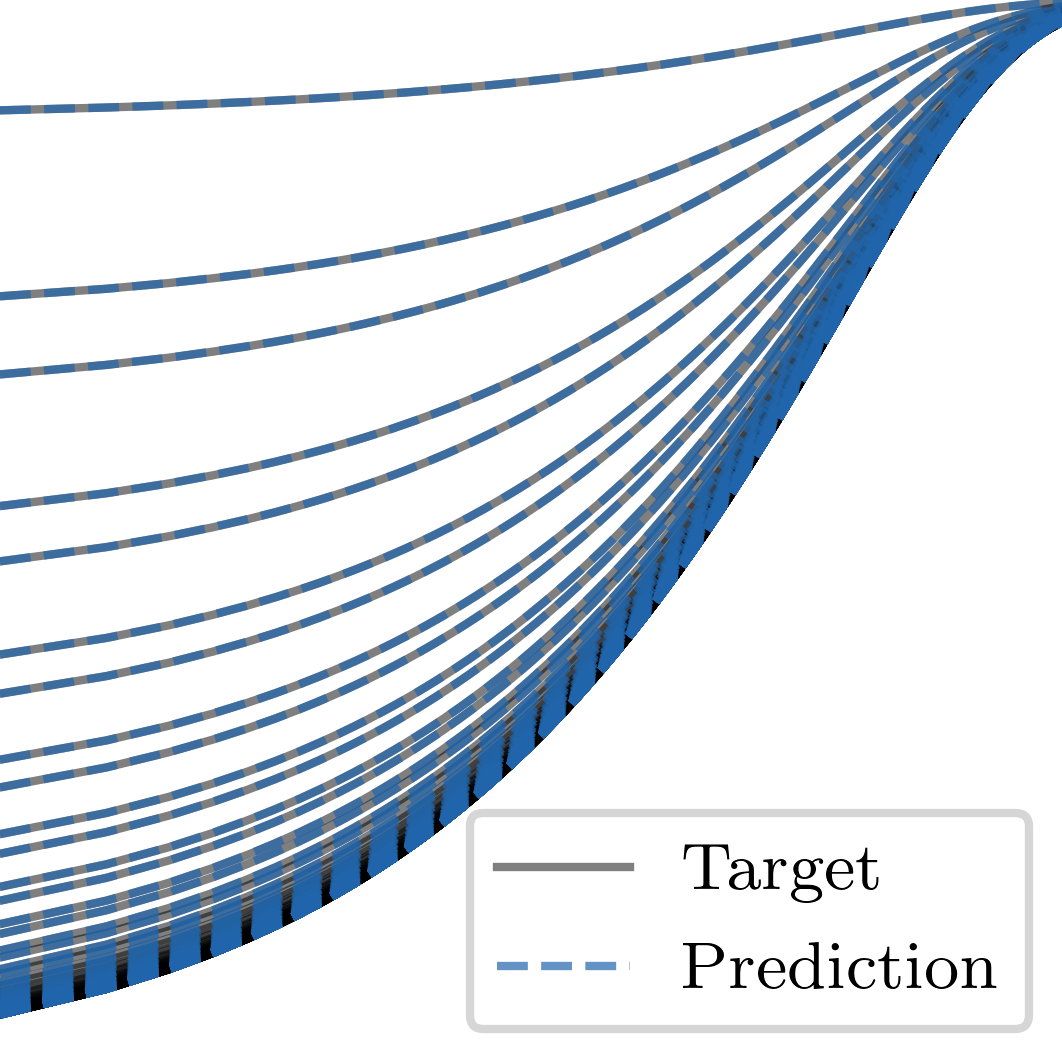};
        \end{loglogaxis}
  \end{tikzpicture}
  \subcaption{Single-step ahead (train)}
  \label{fig:4a}
\end{minipage}%
\begin{minipage}[b]{0.5\textwidth}
\centering
    \begin{tikzpicture}
        \begin{loglogaxis}[
            xminorticks=true,
            xmin=0.1, xmax=10e1,
            yminorticks=true,
            ymin=10, ymax=10e5,
            xlabel = Time (s),
            xlabel style={font=\fontsize{8}{144}\selectfont\color{white!15!black}},
            ylabel = $\overline{p}_1$ (Pa),
            ylabel style={font=\fontsize{8}{144}\selectfont\color{white!15!black}},
            ticklabel style={font=\fontsize{8}{144}},
            width=4.5cm, height=4.5cm,
            scale only axis,
            enlargelimits=false,
            axis on top]
            \addplot graphics[xmin=0.1, xmax=10e1, ymin=10, ymax=10e5] {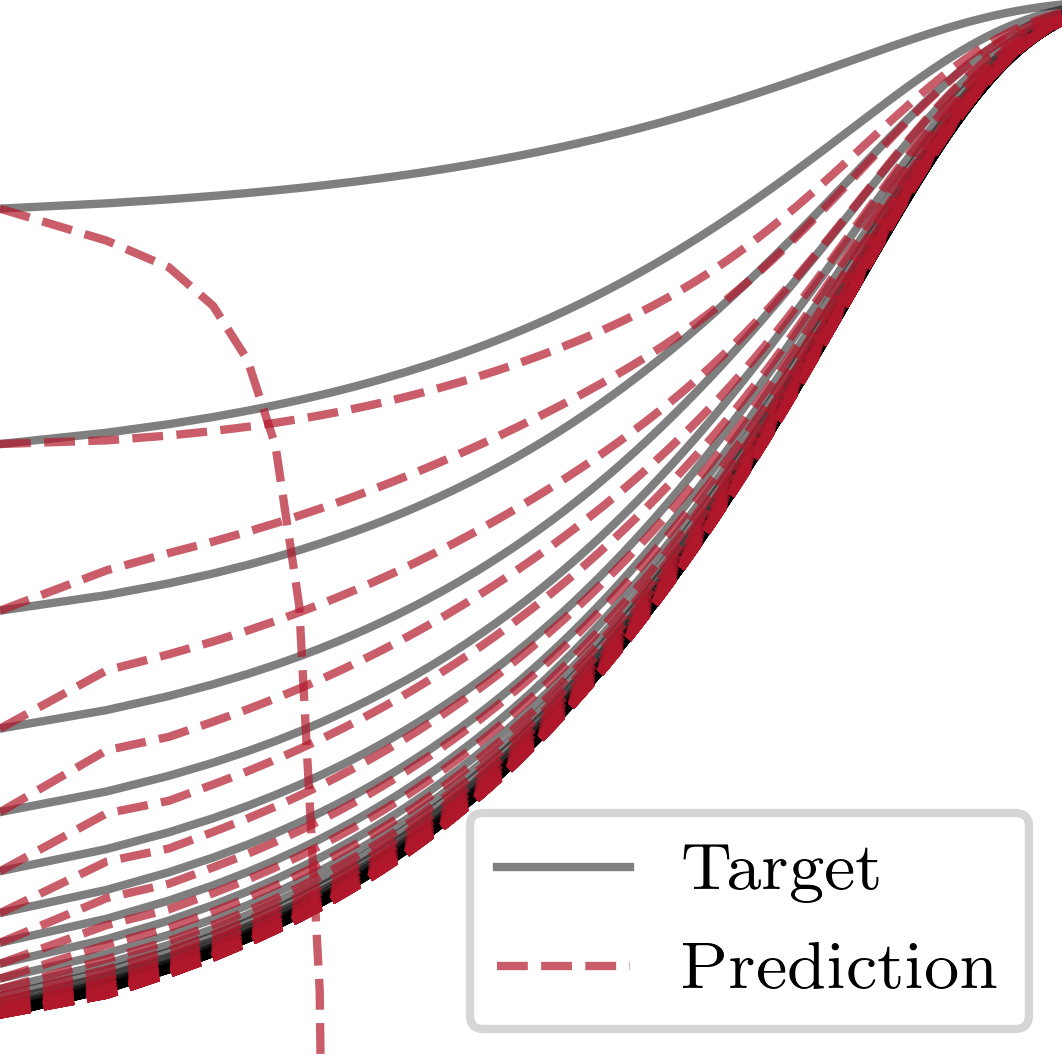};
        \end{loglogaxis}
  \end{tikzpicture}
  \subcaption{Multi-step ahead (test)} \label{fig:4b}
\end{minipage}
\caption{Results for the unregularised polynomial regression: Single-step ahead training (a), and multi-step ahead testing (b).} \label{fig:4}
\end{figure}

\cref{fig:4} shows the train and test results using the unregularised polynomial regression algorithm. From \cref{fig:4a}, we can see there is a good match when training as a single-step ahead prediction problem. The quality of this match is corroborated by the high evaluation metric (and low RMSE respectively) shown in \cref{tab:1}. However, during testing, the prediction quality deteriorates, particularly for sequences associated with high initial pressures (\cref{fig:4b}). Further, for the series corresponding to the highest initial pressure, we observe unbounded error growth with the prediction rapidly going negative. Accordingly, this latter unbounded prediction leads to the missing entries for the error metrics for the test problem shown in \cref{tab:1}. We hypothesised errors in the multi-step ahead case with this model to be caused by the multicollineriaty existing between the transformed features, as described in Section 6.2. Accordingly, to address this problem we introduced a regularised PR by way of a penalisation term on the loss. 

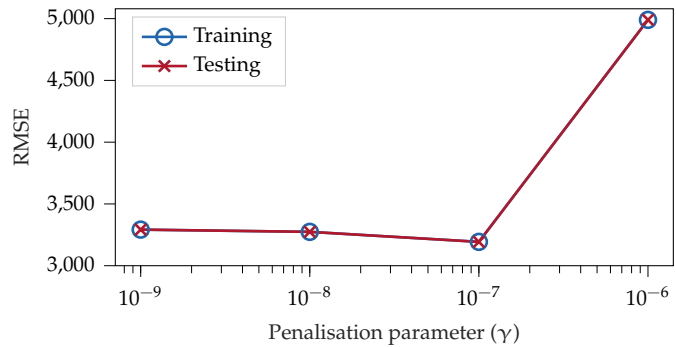
\begin{figure}[h!]
\centering
\setlength\figureheight{5cm}
\setlength\figurewidth{9cm} 

\begin{tikzpicture}

\definecolor{color0}{rgb}{0.1294,0.4,0.6745}
\definecolor{color1}{rgb}{0.698,0.0941,0.1686}

\begin{axis}[
height=\figureheight,
legend cell align={left},
legend style={
  font=\fontsize{8}{144}\selectfont,
  fill opacity=0.8,
  draw opacity=1,
  text opacity=1,
  at={(0.03,0.97)},
  anchor=north west,
  draw=white!80!black
},
log basis x={10},
tick align=outside,
tick pos=left,
ticklabel style={font=\fontsize{8}{144}},
width=\figurewidth,
xlabel style={font=\fontsize{8}{144}\selectfont\color{white!15!black}},
xlabel={Penalisation parameter ($\gamma$)},
xmin=7.07945784384137e-10, xmax=1.41253754462276e-06,
xmode=log,
xtick style={color=black},
ylabel style={font=\fontsize{8}{144}\selectfont\color{white!15!black}},
ylabel={RMSE},
ymin=3000, ymax=5079.85,
ytick style={color=black}
]
\addplot [color0, line width=1.0pt, mark=o, mark size=3, mark options={solid,fill opacity=0}]
table {%
1e-06 4990
1e-07 3193
1e-08 3274
1e-09 3292
};
\addlegendentry{Training}
\addplot [color1, line width=1.0pt, mark=x, mark size=3, mark options={solid}]
table {%
1e-06 4990
1e-07 3193
1e-08 3274
1e-09 3292
};
\addlegendentry{Testing}
\end{axis}

\end{tikzpicture}
\caption{Multi-step ahead training and testing RMSE for different penalisation parameter values in the regularised polynomial regression.} \label{fig:5}
\end{figure}

\cref{fig:5} shows the penalisation parameter tuning for the regularised PR. Accordingly, we observe as $\gamma$ increases from $\gamma=1\times10^{-9}$ to $\gamma=1\times10^{-7}$ the RMSE on both the training and testing sets (treated as both multi-step ahead problems for this demonstration) decreases. However, from $\gamma=1\times10^{-7}$ to $\gamma=1\times10^{-6}$, we can see a sharp increase in RMSE. Indeed, past $\gamma=1\times10^{-6}$ our RMSE becomes unbounded as the accumulating errors become significant. The results from \cref{fig:5} support our use of $\gamma=1\times10^{-7}$ for the weight penalisation. 

\begin{figure}[h!]
\centering
\begin{minipage}[b]{0.5\textwidth}
\centering
    \begin{tikzpicture}
        \begin{loglogaxis}[
            xminorticks=true,
            xmin=0.1, xmax=10e1,
            yminorticks=true,
            ymin=10, ymax=10e5,
            xlabel = Time (s),
            xlabel style={font=\fontsize{8}{144}\selectfont\color{white!15!black}},
            ylabel = $\overline{p}_1$ (Pa),
            ylabel style={font=\fontsize{8}{144}\selectfont\color{white!15!black}},
            ticklabel style={font=\fontsize{8}{144}},
            width=4.5cm, height=4.5cm,
            scale only axis,
            enlargelimits=false,
            axis on top]
            \addplot graphics[xmin=0.1, xmax=10e1, ymin=10, ymax=10e5] {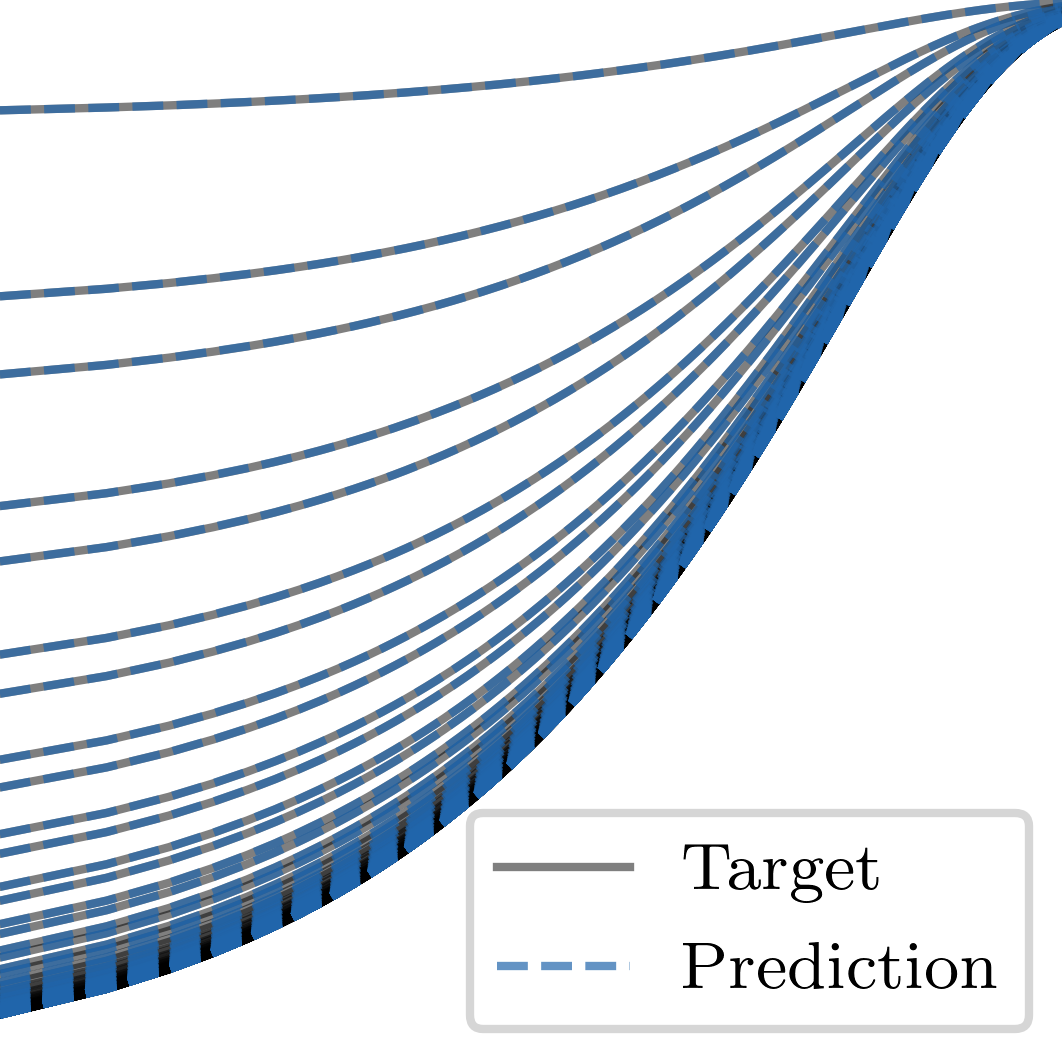};
        \end{loglogaxis}
  \end{tikzpicture}
  \subcaption{Single-step ahead (train)}
  \label{fig:6a}
\end{minipage}%
\begin{minipage}[b]{0.5\textwidth}
\centering
    \begin{tikzpicture}
        \begin{loglogaxis}[
            xminorticks=true,
            xmin=0.1, xmax=10e1,
            yminorticks=true,
            ymin=10, ymax=10e5,
            xlabel = Time (s),
            xlabel style={font=\fontsize{8}{144}\selectfont\color{white!15!black}},
            ylabel = $\overline{p}_1$ (Pa),
            ylabel style={font=\fontsize{8}{144}\selectfont\color{white!15!black}},
            ticklabel style={font=\fontsize{8}{144}},
            width=4.5cm, height=4.5cm,
            scale only axis,
            enlargelimits=false,
            axis on top]
            \addplot graphics[xmin=0.1, xmax=10e1, ymin=10, ymax=10e5] {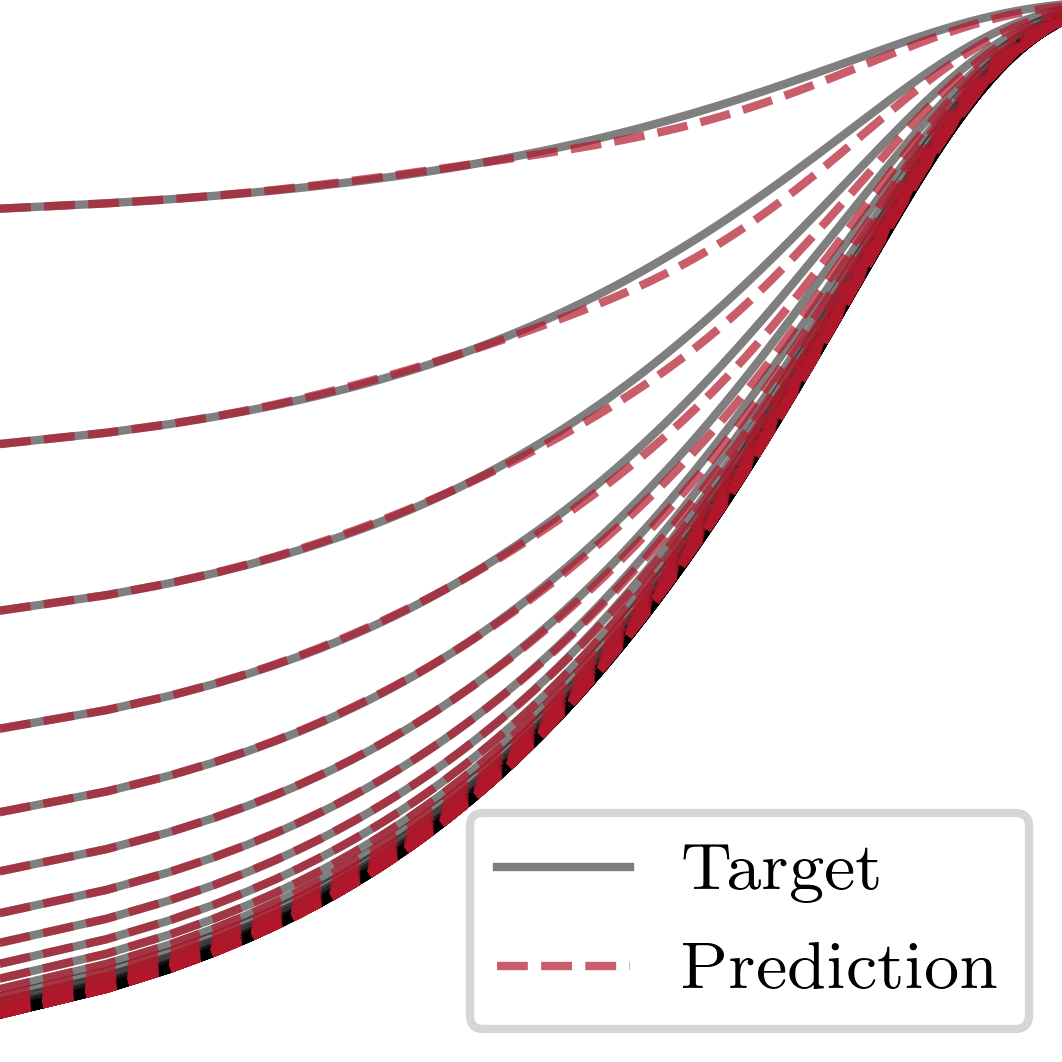};
        \end{loglogaxis}
  \end{tikzpicture}
  \subcaption{Multi-step ahead (test)} \label{fig:6b}
\end{minipage}
\caption{Results for the regularised polynomial regression: Single-step ahead training (a), and multi-step ahead testing (b).} \label{fig:6}
\end{figure}

Results for the regularised PR are shown in \cref{fig:6} and \cref{tab:1}. The training results shown in \cref{fig:6a} are virtually identical to those shown in \cref{fig:4a} for the unregularised PR case (\cref{tab:1}). However, we can observe measurable differences on the multi-steap ahead test set between the two polynomial regression models. From \cref{fig:6b}, we see the regularised PR predictions are qualitatively closer to the targets, contrary to \cref{fig:4b}. Further, we do not observe the unbounded predictions shown in the latter figure. Looking at the error metrics for the regularised PR model in \cref{tab:1}, we see the RMSE is significantly higher for the testing set. Such differences are expected given the accumulation of errors, with the higher test error likely due to the small discrepancies observed at late times in \cref{fig:6b}. Nonetheless, the results using the simple (regularised) polynomial regression model are both qualitatively and quantitatively reasonable, serving as a good benchmark for the more complex methods to follow.  

\begin{figure}[h]
\centering
\begin{minipage}[b]{0.5\textwidth}
\centering
    \begin{tikzpicture}
        \begin{loglogaxis}[
            xminorticks=true,
            xmin=0.1, xmax=10e1,
            yminorticks=true,
            ymin=10, ymax=10e5,
            xlabel = Time (s),
            xlabel style={font=\fontsize{8}{144}\selectfont\color{white!15!black}},
            ylabel = $\overline{p}_1$ (Pa),
            ylabel style={font=\fontsize{8}{144}\selectfont\color{white!15!black}},
            ticklabel style={font=\fontsize{8}{144}},
            width=4.5cm, height=4.5cm,
            scale only axis,
            enlargelimits=false,
            axis on top]
            \addplot graphics[xmin=0.1, xmax=10e1, ymin=10, ymax=10e5] {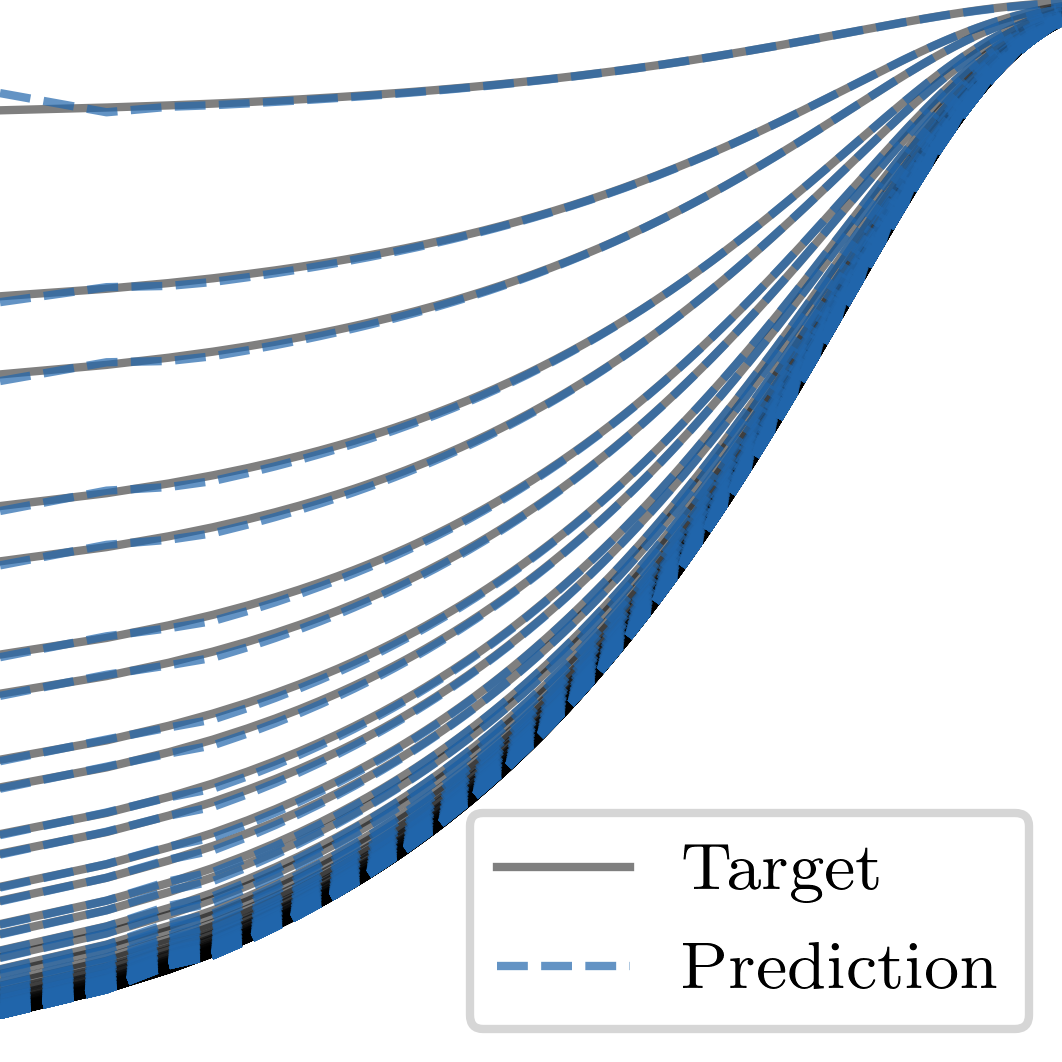};
        \end{loglogaxis}
  \end{tikzpicture}
  \subcaption{Single-step ahead (train)}
  \label{fig:7a}
\end{minipage}%
\begin{minipage}[b]{0.5\textwidth}
\centering
    \begin{tikzpicture}
        \begin{loglogaxis}[
            xminorticks=true,
            xmin=0.1, xmax=10e1,
            yminorticks=true,
            ymin=10, ymax=10e5,
            xlabel = Time (s),
            xlabel style={font=\fontsize{8}{144}\selectfont\color{white!15!black}},
            ylabel = $\overline{p}_1$ (Pa),
            ylabel style={font=\fontsize{8}{144}\selectfont\color{white!15!black}},
            ticklabel style={font=\fontsize{8}{144}},
            width=4.5cm, height=4.5cm,
            scale only axis,
            enlargelimits=false,
            axis on top]
            \addplot graphics[xmin=0.1, xmax=10e1, ymin=10, ymax=10e5] {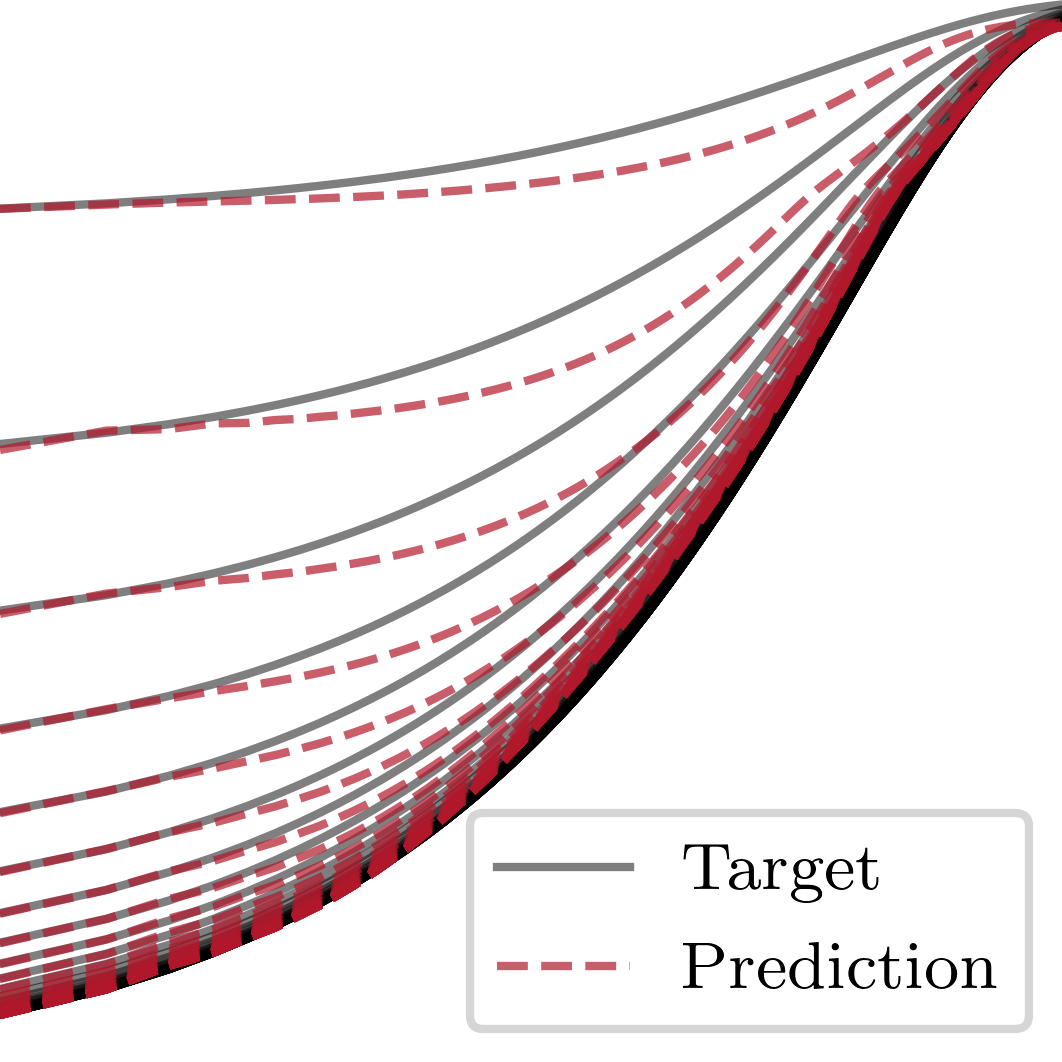};
        \end{loglogaxis}
  \end{tikzpicture}
  \subcaption{Multi-step ahead (test)} \label{fig:7b}
\end{minipage}
\caption{Results for the fully connected neural network: Single-step ahead training (a), and multi-step ahead testing (b).} \label{fig:7}
\end{figure}

Next, \cref{fig:7} and \cref{tab:1} show the train and test results using the FC-NN. From \cref{fig:7a}, we see a good match between the training predictions and targets. However, comparison of the quantitative training results in \cref{tab:1} between the FC-NN and the regularised PR suggests underfitting with the former. From \cref{fig:7b}, we can observe good qualitative matches between the predictions and targets over early-time. However, from middle to late times the predictions diverge from the targets, suggesting the effects of accumulating errors on the multi-step predictions over these time periods. The resulting degradation in performance is reflected in the error metrics in \cref{tab:1} for the FC-NN on the testing case. We observe the diverging prediction behaviour most prominently over the middle to late time periods for sequences corresponding to middle to high initial pressures. One possible explanation for this observation is the data sparsity of these sequences relative to the concentration of sequences at low initial pressures. Accordingly, learning is biased towards fitting those sequences corresponding to low initial pressures. To address this behaviour one could generate more sequence data for middle to high initial pressures. Further, tackling the underfitting problem is non-trivial and in our case was not simply addressed by using wider and deeper network architectures. Interestingly however, despite performing worse than the simple PR model, the FC-NN does not suffer from the same multicollinearity problems as the latter. Such a feature is useful for future investigations involving neural networks.

\begin{table}[h]
\centering
\begin{tabular} {c | c | c | c | c}
Algorithm & $(\text{RMSE)}_\text{train}$ & $(1-\text{NRMSE)}_\text{train}$ & $(\text{RMSE)}_\text{test}$ & $(1-\text{NRMSE)}_\text{test}$ \\ 
\hline \hline 
UPR & 3.43 & 0.999 & -- & -- \\ 
RPR & 3.57 & 0.999 & 3200 & 0.983 \\ 
FC-NN & 266 & 0.998 & 13700 & 0.926 \\
\end{tabular}
\caption{Train and test evaluations for the different learning algorithms. UPR, RPR, and FC-NN denote the unregularised and regularised polynomial regressors, and fully connected neural network respectively. The missing entry for UPR on the test is due to unbounded predictions.} \label{tab:1}
\end{table}

In summary, we have shown training and testing results for the time-dependent problem using a several data-driven modelling strategies. We have seen that complexity does not always result in superior performance, with the simple regularised polynomial regression model outperforming the neural network based approach on both training and testing problems. Indeed, as part of this work, we also experimented with fitting even more complex neural networks, so-called encoder-decoder long short-term memory networks \footnote{See \href{https://github.com/mashworth11/ML-MM}{{\fontfamily{qcr}\selectfont https://github.com/mashworth11/ML-MM}} for further details.}. However, we found these recurrent neural networks challenging to fit in practice, with the results falling short of the performance of the polynomial regression models. For more complex problems, the additional effort required to fit such models may be warranted.  For this work, however, given the results shown in this section, we use the regularised PR model for the next stage of the framework application.   

\section{Application of the framework: Model coupling}
Here we couple our trained surrogate constitutive model to a macroscopic physics-based model. We test the subsequent hybrid ML-physics model on a variety of cases against a traditional physics-based model that uses the linear transfer, and a microscale model.

\subsection{Hybrid ML-physics model}
Our physics-based numerical model is the discretised counterpart to \crefrange{eqn:35}{eqn:36}. Time discretisation is achieved using the backward Euler method, as per \cref{eqn:44}. Discretisation for the pressure gradient is done using the finite-volume method by way of the two-point flux approximation. The latter is implemented as standard within MRST. Given the propensity for nonlinear phenomena in subsurface flow (e.g. due to compressibility) it is common to use nonlinear solvers.  Accordingly,
we give \crefrange{eqn:35}{eqn:36} in discrete block matrix form following application of Newton's method as 
\begin{equation}
    \begin{bmatrix}
    \bm{F}_1 & \bm{G}_1 \\
    \bm{G}_2 & \bm{F}_2 
    \end{bmatrix}^{(k)}
    \begin{bmatrix}
        \delta\hat{\bm{p}}_1 \\
        \delta\hat{\bm{p}}_2
    \end{bmatrix}^{t+1, (k+1)}
    =
    -\begin{bmatrix}
        \bm{R}_{1} \\
        \bm{R}_{2} 
    \end{bmatrix}^{t+1, (k)}, \label{eqn:56}
\end{equation}
where $\delta$ and $k$ now denote the change in solution and current iteration levels respectively. The flow matrix in the Jacobian is $\bm{F}_\alpha=\bm{Q}_\alpha+\Delta t \bm{T}_\alpha$, where $\bm{Q}_\alpha$ is the compressibility matrix and $\bm{T}_\alpha$ is the transmissibility matrix (see \citealt{Lie2019} for details). For the matrix material in the dual-porosity setting, the transmissibility matrix has zero entries. Matrix $\bm{G}_\alpha$ is the coupling matrix arising from an inter-porosity flow expression such as the linear approximation in \cref{eqn:42} (see \cite{Ashworth2020} for details). Note, terms arising from the linear transfer approximation will also appear in the flow matrices. The solution vectors are $\hat{\bm{p}}_\alpha=[\hat{p}_{\alpha}^1,...,\hat{p}_{\alpha}^{n_{cell}}]^\top$, where $\hat{p}_\alpha$ denotes the cell-wise pressure solution for continuum $\alpha$. Notation $n_{cell}$ is then the number of grid cells (control volumes). Next, $\bm{R}^{t+1, (k)}_\alpha=[R_{\alpha}^1,...,R_{\alpha}^{n_{cell}}]^\top$ are the residual vectors for continuum $\alpha$ per grid cell. Intuitively, $R^{t+1, (k)}_\alpha$ provides the difference between the rate of change of fluid mass within a cell volume to that generated by sources and/or fluid movement through its boundaries at iteration level $k$. Finally, we inject our machine learning model into \cref{eqn:56} through an explicit approach. As a result, following a Newton iteration the coupling matrices $\bm{G}_\alpha$ in \cref{eqn:56} are zero entries, whilst the machine learning-based $\overline{r}_1$ appears in the residuals. 

\subsection{Tests}
To test the hybrid ML-physics model we consider several realisations of the 2D diffusion problem described in Section 4. For these realisations, we test our framework on initial conditions and time lengths not used for training our model. Accordingly, we compare the hybrid approach against a microscale model and a physics-based model that uses the first-order mass transfer constitutive relation. For brevity in the remainder of this section, we refer to the latter method as the `traditional' approach to dual-continuum modelling. In addition to the 2D problem, we also consider an application of the framework to a geological model. For this case, we only compare the hybrid and traditional dual-porosity approaches. 

For the 2D problem test cases we discretise the DP problem using a single control volume. For the microscale model we resolve the matrix using 40$\times$40 grid cells. Within the DP model we initialise the fracture pressure as $\overline{p}^0_2=1 \text{ MPa}$. Further, to mimic the fracture boundary within the DP model, we set fracture permeability to $10 \text{ d}$. As a result, the fracture pressure equilibrates almost instantaneously with the external boundary in response to changes induced by inter-continuum flow. The remaining properties are as described in Section 5. Specifically, $\overline{\phi}^0_2 = 1\times10^{-3}$, $c_2 = c_1 = 1.4 \text{ GPa}^{-1}$, $v_1=0.999$, $\phi^0_1 = 0.2$, $\mu = 1 \text{ cp}$, $N_f=1$, $\ell=1 \text{ m}$, $\rho^0=1000 \text{ kgm}^{-3}$ and $\mathsf{k}_1=\mathsf{k}'=1 \text{ md}$ . With respect to $\overline{\phi}^0_2$ and $c_2$, these are chosen somewhat arbitrarily since we do not consider fracture dynamics for the problems herein. Lastly, we make use of a discrete equivalent to \cref{eqn:2} to compare the (averaged) microscopic matrix pressure field with its dual-continuum counterpart. 

For the geological model, we use a grid taken from the \cite{OPM}. Accordingly, each control volume now corresponds to a 3D domain involving three orthogonal fracture sets leading to a 3D diffusion problem within the matrix. As a result, the analytical solution used to generate data is a higher dimensional version of \cref{eqn:39}. Further details of this solution, and the parameterisations used for this test, are described as part of the test case descriptions introduced below.

\subsubsection*{Case 1: $p^0_1 \in [1, 1\times10^6] \text{ Pa}$}
For the first set of test cases we consider two initial pressures not used when constructing $\mathcal{D}$, but coming from $[1 , 1\times10^6] \text{ Pa}$. Specifically, we sample the initial pressures such that $p^{0,\text{I}}_1<10 \text{ Pa}$ and $0.1 \text{ MPa}<p^{0,\text{II}}_1$, where superscripts $\text{I}$ and $\text{II}$ correspond to the two tests respectively. The boundary pressure is fixed as $\overline{p}_2 = 1 \text{ MPa}$. The parameters for this 2D problem are as described above.

\subsubsection*{Case 2: $T \in \{1, 10\} \text{ s}$}
For the second set of tests we consider shorter sequences for training and use the resulting models for testing on the full sequence. Specifically, we consider datasets up to $T=1 \text{ s}$ and $T=10 \text{ s}$ for training, instead of a `full' dataset corresponding to $T=100 \text{ s}$. The motivation behind this test case is to see how the models perform when trained under limited data and used beyond the training data range. Accordingly, extrapolation outside of the training conditions is now caused by the length of the testing sequence length relative to the training sequence length, as well as due to the multi-step ahead approach. The parameters for this 2D problem are as described above, with $p^0_1=1 \text{ Pa}$.

\subsubsection*{Case 3: Application to a geological model}
For the 3D problem, \cref{eqn:39} is extended as (\citealt{Lim1995, Zhou2017}),
\begin{align}
    \overline{p}_1 = p^0_1 + (&\overline{p}_2-p^0_1)\Bigg\{1-\sum_{n=0}^{\infty}\sum_{m=0}^{\infty}\sum_{o=0}^{\infty}\left(\frac{8}{\pi^2}\right)^3\frac{1}{(2n+1)^2(2m+1)^2(2o+1)^2} \nonumber \\
    &\text{exp}\left(-\frac{\pi^2\mathsf{k}_1t}{\phi^0_1c_1 \mu\ell^2}\left[(2n+1)^2+(2m+1)^2+(2o+1)^2\right]\right)\Bigg\} \label{eqn:57}
\end{align}
where we parameterise \cref{eqn:57} using $\overline{p}_2 = 1 \text{ MPa}$, $c_1 = 3 \text{ GPa}^{-1}$, $\phi^0_1 = 0.2$, $\mu = 5 \text{ cp}$, $\ell=1 \text{ m}$ and $\mathsf{k}_1=1\times10^{-4} \text{ md}$. As before, we take 120 different $p^0_1$ sampled from the interval $[1 , 1\times10^6] \text{ Pa}$. Further, we generate series starting from $t=0$ and finishing at $T=100 \text{ hours}$, with a uniform stepping interval of $\Delta t=1 \text{ hour}$. Finally, we train and test on the resulting data using an autoregressive regularised polynomial regression model, which is subsequently injected into our physics-based simulator.

For our geological DP model, the boundaries are specified as pressure boundaries set at $1 \text{ MPa}$. Next, the (macroscopic) fracture permeability is set at $10 \text{ d}$. Accordingly, the high fracture permeability ensures our numerical test problem is consistent with our learning problem in terms of pressure conditions for the fracture. We initialise the matrix as $p^0_1=5 \text{ Pa}$. To complete the DP model we assign $\overline{\phi}^0_2 = 1\times10^{-3}$, $c_2 = 3 \text{ GPa}^{-1}$ and $\rho^0=850 \text{ kgm}^{-3}$. 
We compare results over the geological model between the hybrid ML-physics model and the physics-based model using the linear transfer. For the latter, the transfer function is parameterised using $N_f=3$ and $\mathsf{k}'=1\times10^{-4} \text{ md}$. Finally, despite only training for a series corresponding to $100 \text{ hours}$ we run each simulation for a $1000 \text{ hours}$, thus extrapolating outside of the training conditions.

\subsection{Results}
Here we present the matrix pressure evolution results for the test cases described above. 

\subsubsection*{Case 1: $p^0_1 \in [1 , 1\times10^6] \text{ Pa}$}
Results for the first test case are shown in \cref{fig:8}. In both cases we see pressure increasing with time as the matrix equilibrates with the boundary pressure. We can see for both tests there is a measurable discrepancy between the DP model using a linear transfer and the equivalent microscale description. This discrepancy is particularly pronounced for test I (\cref{fig:8a}). For this test case, the linear transfer is a particularly poor early-time model when the difference between continuum pressures is several orders of magnitude. In contrast, the DP approach with the ML transfer produces high quality matches for both test cases without the computational expense taken to run the microscale models. In \cref{fig:6b}, we observed small errors in the middle to late-time regularised PR predictions for test series corresponding to high initial pressures. However, \cref{fig:8b} shows despite these discrepancies, the DP model using the data-driven transfer still outperforms a linear transfer-based DP model with respect to the microscale solution. 

Lastly, for both tests we also provide the analytical solutions (\cref{fig:8}). The benefit of this addition is clearer in \cref{fig:8a}. From \cref{fig:8a} we observe a small discrepancy between the microscale model and the analytical solution for the first few time steps. This discrepancy arises due to the time step discretisation error in the microscale model. Remarkably, we do not see this discrepancy between the ML-based approach and the analytical solution, since the learnt model comes from data unaffected by these discretisation errors. This subtle result corroborates the aim of multiscale approaches. Specifically, to combine the accuracy of microscopic representations with the practicality of macroscopic models.

\begin{figure}[h]
\centering
\begin{minipage}[b]{0.5\textwidth}
\centering
    \setlength\figureheight{4.75cm}
    \setlength\figurewidth{4.75cm}  
    \definecolor{mycolor1}{rgb}{0.12941,0.40000,0.67451}%
    \definecolor{mycolor2}{rgb}{0.69804,0.09412,0.16863}%
    \definecolor{mycolor3}{rgb}{0.46600,0.67400,0.18800}%
%
    \subcaption{$p^{0,\text{I}}_1 = \text{3 Pa}$}
    \label{fig:8a}
\end{minipage}%
\begin{minipage}[b]{0.5\textwidth}
\centering
    \setlength\figureheight{4.75cm}
    \setlength\figurewidth{4.75cm}  
    \definecolor{mycolor1}{rgb}{0.12941,0.40000,0.67451}%
    \definecolor{mycolor2}{rgb}{0.69804,0.09412,0.16863}%
    \definecolor{mycolor3}{rgb}{0.46600,0.67400,0.18800}%
    %
%
    \subcaption{$p^{0,\text{II}}_1 = 0.8 \times10^6 \text{ Pa}$} 
    \label{fig:8b}
\end{minipage}%
\caption{Average matrix pressure evolutions for test case 1:  $p^{0,\text{I}}_1 = 3 \text{ Pa}$ (a) and $p^{0,\text{II}}_1 = 0.8\times10^6 \text{ Pa}$ (b). Notations `DP-L' and `DP-ML' denote the dual-porosity model equipped with the linear and machine learning constitutive transfer models respectively.} \label{fig:8}
\end{figure}

\subsubsection*{Case 2: $T \in \{1, 10\} \text{ s}$}
\begin{figure}[h]
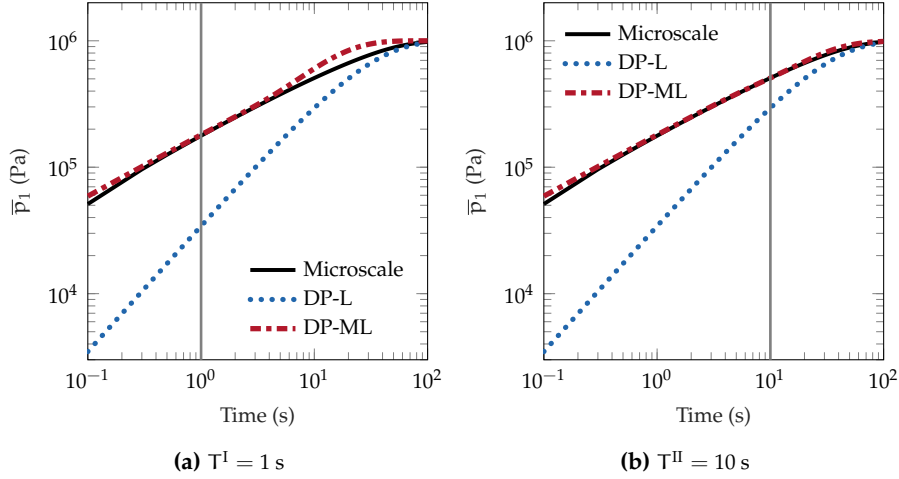

\centering
\begin{minipage}[b]{0.5\textwidth}
\centering
    \setlength\figureheight{4.75cm}
    \setlength\figurewidth{4.75cm}  
    \definecolor{mycolor1}{rgb}{0.12941,0.40000,0.67451}%
    \definecolor{mycolor2}{rgb}{0.69804,0.09412,0.16863}%
%
    \subcaption{$T^{\text{I}} = 1 \text{ s}$}
    \label{fig:9a}
\end{minipage}%
\begin{minipage}[b]{0.5\textwidth}
\centering
    \setlength\figureheight{4.75cm}
    \setlength\figurewidth{4.75cm}  
    \definecolor{mycolor1}{rgb}{0.12941,0.40000,0.67451}%
    \definecolor{mycolor2}{rgb}{0.69804,0.09412,0.16863}%
    %
%
    \subcaption{$T^{\text{II}} = 10 \text{ s}$} 
    \label{fig:9b}
\end{minipage}%
\caption{Average matrix pressure evolutions for test case 2:  $T^{\text{I}} = 1 \text{ s}$ (a) and $T^{\text{II}} = 10 \text{ s}$ (b). Notations `DP-L' and `DP-ML' denote the dual-porosity model equipped with the linear and machine learning constitutive transfer models respectively.} \label{fig:9}
\end{figure}

\cref{fig:9} shows the results for the different training and testing sequence length test cases. \cref{fig:9a} shows the test case when using up to $T=1 \text{ s}$ of data (as denoted by the grey vertical line). In this figure, we observe a good qualitative match relative to the microscale model using the hybrid approach over the first $1 \text{ s}$. Further, even for time steps within a few seconds after the $1 \text{ s}$ training limit, the match between the hybrid and microscale models is good. However, from middle to late times, \cref{fig:9a} shows we see the hybrid model diverging from the microscale model solution. However, at late times the model converges back to the microscale solution. In \cref{fig:9b}, the sequence limit is now at $T=10 \text{ s}$. As one would expect, more data used for our surrogate model results in improved hybrid model performance across the sequence length. Further, in \cref{fig:9b} there is a noticeably smaller divergence between the hybrid and microscale model results following the $10 \text{ s}$ limit, compared to that shown in \cref{fig:9a} following the $1 \text{ s}$ limit. This result is consistent with the general trend between surrogate model accuracy and data volume, leading to a lower accumulation of errors. Nonetheless, even with a surrogate constutive model trained using $T=1 \text{ s}$ of sequence data, the hybrid model still performs well over the whole simulation time. 

From the first two test cases, we have seen a traditional dual-porosity model approach gives the worst results at early times. At these times, there is a significant different between matrix and boundary pressures. However, at later times, the linear transfer relation is a reasonable approximation of inter-continuum mass exchange. Accordingly, it is wasteful to use a surrogate constitutive model in regions where a simple constitutive model performs well, given the data requirements to train the former. An interesting development, therefore, would be a hybrid model that can use different constitutive models in different regions (e.g. nonlinear vs linear) of a problem space (\citealt{March2016}). 

\subsubsection*{Case 3: Application to a geological model}
\begin{figure}[h!]
\centering
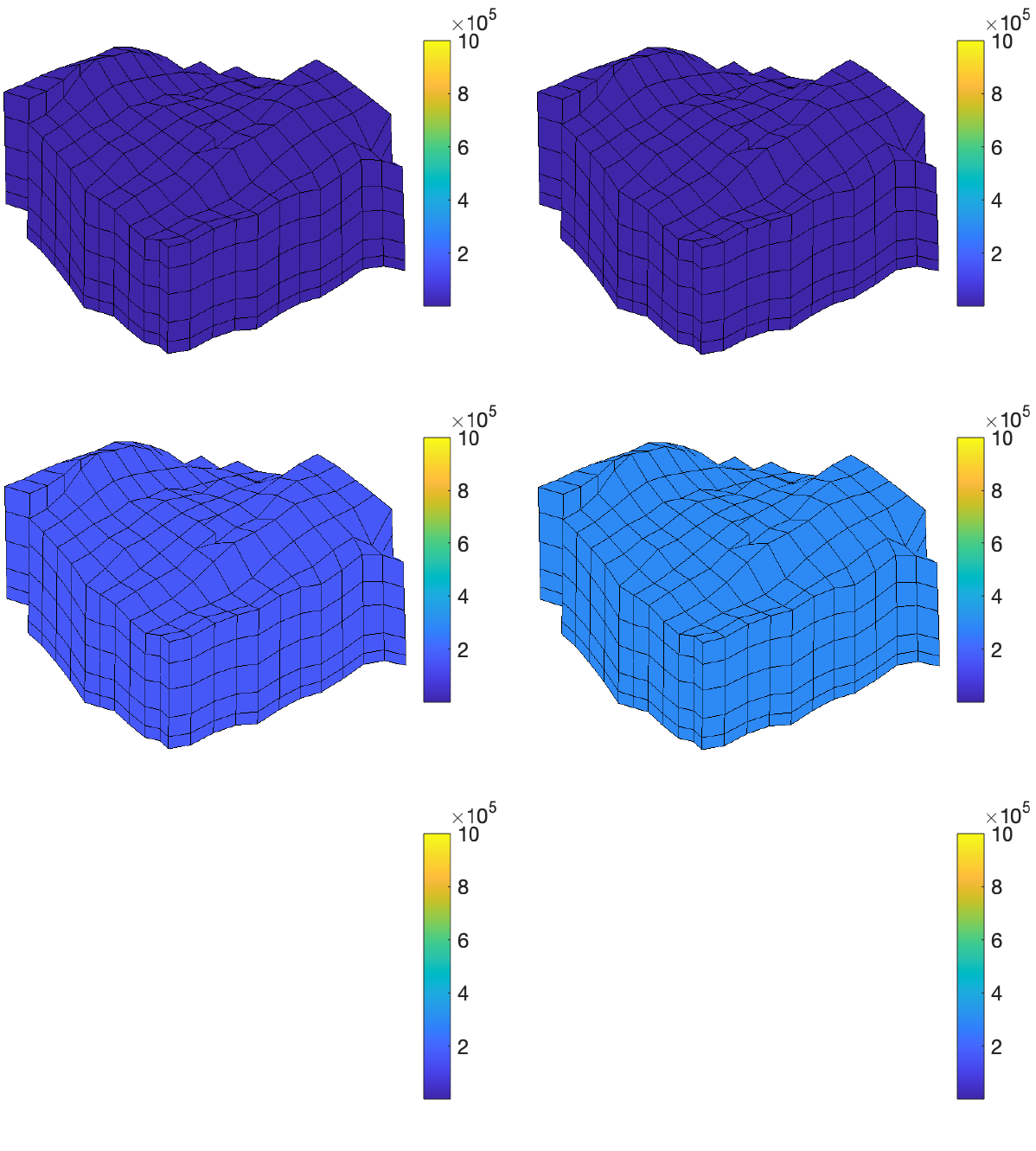
\caption{Matrix pressure evolutions on a geological model using the linear transfer model.}
\label{fig:10}
\end{figure}

\begin{figure}[h!]
\centering
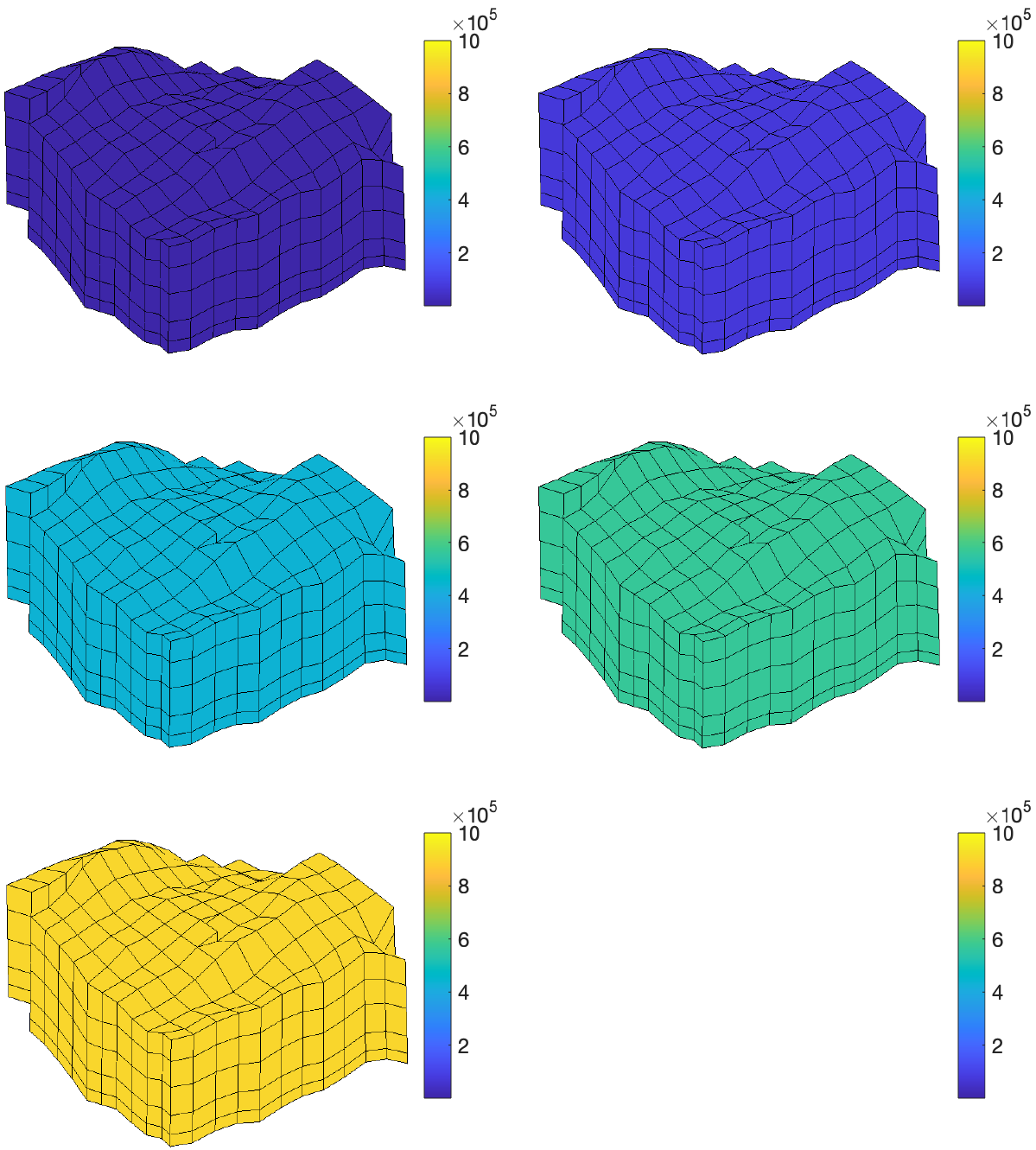
\caption{Matrix pressure evolutions on a geological model using the machine learning-based transfer model.}
\label{fig:11}
\end{figure}

\crefrange{fig:10}{fig:11} show the results for the geological model test case using the traditional and ML-physics model respectively. In both figures we can see matrix pressure rising over time in response to diffusion driven equilibration with the fractures. However, we can see significant differences between the two modelling approaches following the initial conditions up until equilibrium conditions. Comparing \crefrange{fig:10}{fig:11} at t = 1 hour, we can see the matrix pressures are noticeably higher when using the hybrid ML-physics approach. This observation is consistent with the results in the previous tests, which showed the matrix pressures to be significantly underestimated when using a linear approximation. Accordingly, matrix fluxes are underpredicted when using this linear transfer approach. Such errors could be significant in geological settings such as groundwater remediation (\citealt{Haggerty1995}). Similar differences between the two approaches to those just described are observed at t = 50, 100 hours, with the ML-physics approach predicting higher pressures than the traditional approach. 

When extrapolating outside of the training conditions, the hybrid approach shows physically reasonable results (consistent with the previous time steps) at t = 500 hours. Accordingly, at this time step, matrix pressures coming from the hybrid approach are higher compared to the traditional approach. However, by t = 1000 hours, the two models have converged to the equilibrated pressure state. 

In summary, the results presented for the described tests show the potential for the ML-physics approach for multiscale modelling. With this framework we can capture the accuracy of microscopic representations with the practicality of macroscopic models. We envision a number of interesting extensions to be investigated using the basis of the framework introduced herein.  

\section{Conclusions and future work}
This paper introduces and applies a machine learning-based multiscale constitutive modelling framework. Accordingly, we detailed the key components of the framework as: Homogenisation, data generation, surrogate constitutive model learning, and ML-physics model coupling. In doing, we described various considerations that the user should make at each step. To test the framework we considered a multiscale pressure diffusion problem. Specifically, that of describing inter-continuum mass transfer in double-porosity materials. In practice, it is common to model this process using an overly simple linear constitutive relation. Instead of making any such assumption, we use our framework to build a surrogate constitutive model based on microscale data. In applying the framework we introduced the various machine learning methods used to create this surrogate constitutive model. We account for time-dependence using autoregressive approaches, considering polynomial and neural network-based regressors. We found the former to provide higher accuracy when used as a multi-step ahead predictor, provided regularisation is used to deal with multicollinearity effects. Accordingly, we integrated the resulting data-driven surrogate into a physics-based dual-porosity model, creating a hybrid ML-physics approach. We showed the resulting hybrid approach to give high quality results for a number of test cases compared to a linear transfer-based dual-porosity model, without the computational expense of explicit microscale models.

We hypothesise a number of exciting possibilities for further work when using data coming from more complex microscale scenarios (e.g. dynamic boundary conditions) than the ones considered here. In such cases, data is likely to be obtained numerically. However, numerical data can be expensive, leading to sparse datasets. Accordingly, future developments could involve exploring different sampling and learning strategies such as active learning, theory-driven ML and probablistic approaches to improve learning efficiency, generalisability and to understand uncertainty in multiscale settings. 

\section*{Acknowledgments}
The authors are grateful for the funding provided to them by the Natural Environmental Research Council to carry out this work, and to Anne-Sophie Ruget for advice on figures.

\newpage
\bibliographystyle{agsm}

\end{document}

%% file: Figures/ml_based_2.pdf_tex
\begingroup%
  \makeatletter%
  \providecommand\color[2][]{%
    \errmessage{(Inkscape) Color is used for the text in Inkscape, but the package 'color.sty' is not loaded}%
    \renewcommand\color[2][]{}%
  }%
  \providecommand\transparent[1]{%
    \errmessage{(Inkscape) Transparency is used (non-zero) for the text in Inkscape, but the package 'transparent.sty' is not loaded}%
    \renewcommand\transparent[1]{}%
  }%
  \providecommand\rotatebox[2]{#2}%
  \newcommand*\fsize{\dimexpr\f@size pt\relax}%
  \newcommand*\lineheight[1]{\fontsize{\fsize}{#1\fsize}\selectfont}%
  \ifx\svgwidth\undefined%
    \setlength{\unitlength}{276.85820679bp}%
    \ifx\svgscale\undefined%
      \relax%
    \else%
      \setlength{\unitlength}{\unitlength * \real{\svgscale}}%
    \fi%
  \else%
    \setlength{\unitlength}{\svgwidth}%
  \fi%
  \global\let\svgwidth\undefined%
  \global\let\svgscale\undefined%
  \makeatother%
  \begin{picture}(1,0.93426838)%
    \lineheight{1}%
    \setlength\tabcolsep{0pt}%
    \put(0,0){\includegraphics[width=\unitlength,page=1]{ml_based_2.pdf}}%
    \put(0.69382165,0.58468438){\makebox(0,0)[lt]{\lineheight{1.25}\smash{\begin{tabular}[t]{l}{\small (2) Data generation}\end{tabular}}}}%
    \put(-0.00051851,0.04796889){\makebox(0,0)[lt]{\lineheight{1.25}\smash{\begin{tabular}[t]{l}{\small  (3) Surrogate constitutive}\end{tabular}}}}%
    \put(0.0800653,0.00443383){\makebox(0,0)[lt]{\lineheight{1.25}\smash{\begin{tabular}[t]{l}{\small  model learning}\end{tabular}}}}%
    \put(0.6775376,0.04796889){\makebox(0,0)[lt]{\lineheight{1.25}\smash{\begin{tabular}[t]{l}{\small (4) Model coupling}\end{tabular}}}}%
    \put(0.7025386,0.25908573){\makebox(0,0)[lt]{\lineheight{1.25}\smash{\begin{tabular}[t]{l}$\hat{f}(\mathsfbf{x})\rightarrow$\end{tabular}}}}%
    \put(0.15520189,0.30686905){\makebox(0,0)[lt]{\lineheight{1.25}\smash{\begin{tabular}[t]{l}$f(\mathsfbf{X})$\end{tabular}}}}%
    \put(0.15465657,0.19945018){\makebox(0,0)[lt]{\lineheight{1.25}\smash{\begin{tabular}[t]{l}$\hat{f}(\mathsfbf{X})$\end{tabular}}}}%
    \put(0.17401576,0.25363177){\makebox(0,0)[lt]{\lineheight{1.25}\smash{\begin{tabular}[t]{l}$\approx$\end{tabular}}}}%
    \put(0.49023591,0.29373465){\makebox(0,0)[lt]{\lineheight{1.25}\smash{\begin{tabular}[t]{l}$\hat{f}(\cdot)$\end{tabular}}}}%
    \put(0.4099294,0.51827701){\makebox(0,0)[lt]{\lineheight{1.25}\smash{\begin{tabular}[t]{l}$(\mathsfbf{X}, \mathsfbf{y})$\end{tabular}}}}%
    \put(0.51020517,0.47545076){\makebox(0,0)[lt]{\lineheight{1.25}\smash{\begin{tabular}[t]{l}{\small where} $\mathsfbf{y}=f(\mathsfbf{X})$\end{tabular}}}}%
    \put(0,0){\includegraphics[width=\unitlength,page=2]{ml_based_2.pdf}}%
    \put(0.0535556,0.5846843){\makebox(0,0)[lt]{\lineheight{1.25}\smash{\begin{tabular}[t]{l}{\small (1) Homogenisation}\end{tabular}}}}%
    \put(0.1598818,0.69627371){\makebox(0,0)[lt]{\lineheight{1.25}\smash{\begin{tabular}[t]{l}{\small Macro}\end{tabular}}}}%
    \put(0.16225289,0.86041039){\makebox(0,0)[lt]{\lineheight{1.25}\smash{\begin{tabular}[t]{l}{\small Micro}\end{tabular}}}}%
    \put(0.40950938,0.76264264){\makebox(0,0)[lt]{\lineheight{1.25}\smash{\begin{tabular}[t]{l}{\small Procedures for}\end{tabular}}}}%
    \put(0.40238807,0.68628169){\makebox(0,0)[lt]{\lineheight{1.25}\smash{\begin{tabular}[t]{l}{\small data generation}\end{tabular}}}}%
  \end{picture}%
\endgroup%

%% file: Figures/problem_schematic.pdf_tex
\begingroup%
  \makeatletter%
  \providecommand\color[2][]{%
    \errmessage{(Inkscape) Color is used for the text in Inkscape, but the package 'color.sty' is not loaded}%
    \renewcommand\color[2][]{}%
  }%
  \providecommand\transparent[1]{%
    \errmessage{(Inkscape) Transparency is used (non-zero) for the text in Inkscape, but the package 'transparent.sty' is not loaded}%
    \renewcommand\transparent[1]{}%
  }%
  \providecommand\rotatebox[2]{#2}%
  \newcommand*\fsize{\dimexpr\f@size pt\relax}%
  \newcommand*\lineheight[1]{\fontsize{\fsize}{#1\fsize}\selectfont}%
  \ifx\svgwidth\undefined%
    \setlength{\unitlength}{124.95987532bp}%
    \ifx\svgscale\undefined%
      \relax%
    \else%
      \setlength{\unitlength}{\unitlength * \real{\svgscale}}%
    \fi%
  \else%
    \setlength{\unitlength}{\svgwidth}%
  \fi%
  \global\let\svgwidth\undefined%
  \global\let\svgscale\undefined%
  \makeatother%
  \begin{picture}(1,1.27718064)%
    \lineheight{1}%
    \setlength\tabcolsep{0pt}%
    \put(0,0){\includegraphics[width=\unitlength,page=1]{problem_schematic.pdf}}%
    \put(0.43418785,0.13441301){\makebox(0,0)[lt]{\lineheight{1.25}\smash{\begin{tabular}[t]{l}{\small Matrix}\end{tabular}}}}%
    \put(0,0){\includegraphics[width=\unitlength,page=2]{problem_schematic.pdf}}%
    \put(0.43418785,0.01715107){\makebox(0,0)[lt]{\lineheight{1.25}\smash{\begin{tabular}[t]{l}{\small Fracture}\end{tabular}}}}%
    \put(0,0){\includegraphics[width=\unitlength,page=3]{problem_schematic.pdf}}%
    \put(0.37996148,1.23433541){\makebox(0,0)[lt]{\lineheight{1.25}\smash{\begin{tabular}[t]{l}{\small No flux}\end{tabular}}}}%
    \put(0.37996148,0.28134375){\makebox(0,0)[lt]{\lineheight{1.25}\smash{\begin{tabular}[t]{l}{\small No flux}\end{tabular}}}}%
    \put(0.04284523,0.66322878){\rotatebox{90}{\makebox(0,0)[lt]{\lineheight{1.25}\smash{\begin{tabular}[t]{l}{\small No flux}\end{tabular}}}}}%
    \put(0,0){\includegraphics[width=\unitlength,page=4]{problem_schematic.pdf}}%
    \put(0.95715477,0.90601719){\rotatebox{-90}{\makebox(0,0)[lt]{\lineheight{1.25}\smash{\begin{tabular}[t]{l}{\small No flux}\end{tabular}}}}}%
    \put(0,0){\includegraphics[width=\unitlength,page=5]{problem_schematic.pdf}}%
  \end{picture}%
\endgroup%

%% file: Figures/pressure_step.pdf_tex
\begingroup%
  \makeatletter%
  \providecommand\color[2][]{%
    \errmessage{(Inkscape) Color is used for the text in Inkscape, but the package 'color.sty' is not loaded}%
    \renewcommand\color[2][]{}%
  }%
  \providecommand\transparent[1]{%
    \errmessage{(Inkscape) Transparency is used (non-zero) for the text in Inkscape, but the package 'transparent.sty' is not loaded}%
    \renewcommand\transparent[1]{}%
  }%
  \providecommand\rotatebox[2]{#2}%
  \newcommand*\fsize{\dimexpr\f@size pt\relax}%
  \newcommand*\lineheight[1]{\fontsize{\fsize}{#1\fsize}\selectfont}%
  \ifx\svgwidth\undefined%
    \setlength{\unitlength}{165.43388417bp}%
    \ifx\svgscale\undefined%
      \relax%
    \else%
      \setlength{\unitlength}{\unitlength * \real{\svgscale}}%
    \fi%
  \else%
    \setlength{\unitlength}{\svgwidth}%
  \fi%
  \global\let\svgwidth\undefined%
  \global\let\svgscale\undefined%
  \makeatother%
  \begin{picture}(1,0.95714133)%
    \lineheight{1}%
    \setlength\tabcolsep{0pt}%
    \put(0,0){\includegraphics[width=\unitlength,page=1]{pressure_step.pdf}}%
    \put(0.04263859,0.57239253){\rotatebox{90}{\makebox(0,0)[lt]{\lineheight{1.25}\smash{\begin{tabular}[t]{l} {\small Pressure (Pa)}\end{tabular}}}}}%
    \put(0.31556867,0.85295182){\makebox(0,0)[lt]{\lineheight{1.25}\smash{\begin{tabular}[t]{l}$\overline{p}_2 = 1 \text{ {\small MPa at} } t^{0+}$\end{tabular}}}}%
    \put(-0.00338245,0.30688367){\makebox(0,0)[lt]{\lineheight{1.25}\smash{\begin{tabular}[t]{l}$p^0_1$\end{tabular}}}}%
    \put(0.23003496,0.16868539){\makebox(0,0)[lt]{\lineheight{1.25}\smash{\begin{tabular}[t]{l}$t^{0+}$\end{tabular}}}}%
    \put(0.7103847,0.19451111){\makebox(0,0)[lt]{\lineheight{1.25}\smash{\begin{tabular}[t]{l}{\small Time (s)}\end{tabular}}}}%
    \put(0.65636475,0.36626132){\makebox(0,0)[lt]{\lineheight{1.25}\smash{\begin{tabular}[t]{l}{\small Fracture}\end{tabular}}}}%
    \put(0,0){\includegraphics[width=\unitlength,page=2]{pressure_step.pdf}}%
    \put(0.65799152,0.45087328){\makebox(0,0)[lt]{\lineheight{1.25}\smash{\begin{tabular}[t]{l}{\small Matrix}\end{tabular}}}}%
  \end{picture}%
\endgroup%

%% file: Figures/gg_LT.pdf_tex
\begingroup%
  \makeatletter%
  \providecommand\color[2][]{%
    \errmessage{(Inkscape) Color is used for the text in Inkscape, but the package 'color.sty' is not loaded}%
    \renewcommand\color[2][]{}%
  }%
  \providecommand\transparent[1]{%
    \errmessage{(Inkscape) Transparency is used (non-zero) for the text in Inkscape, but the package 'transparent.sty' is not loaded}%
    \renewcommand\transparent[1]{}%
  }%
  \providecommand\rotatebox[2]{#2}%
  \newcommand*\fsize{\dimexpr\f@size pt\relax}%
  \newcommand*\lineheight[1]{\fontsize{\fsize}{#1\fsize}\selectfont}%
  \ifx\svgwidth\undefined%
    \setlength{\unitlength}{344.744222bp}%
    \ifx\svgscale\undefined%
      \relax%
    \else%
      \setlength{\unitlength}{\unitlength * \real{\svgscale}}%
    \fi%
  \else%
    \setlength{\unitlength}{\svgwidth}%
  \fi%
  \global\let\svgwidth\undefined%
  \global\let\svgscale\undefined%
  \makeatother%
  \begin{picture}(1,1.11133866)%
    \lineheight{1}%
    \setlength\tabcolsep{0pt}%
    \put(0,0){\includegraphics[width=\unitlength,page=1]{gg_LT.pdf}}%
    \put(0.12600875,1.09200017){\makebox(0,0)[lt]{\lineheight{1.25}\smash{\begin{tabular}[t]{l}{\small t = 0 hours}\end{tabular}}}}%
    \put(0.64329316,1.09391948){\makebox(0,0)[lt]{\lineheight{1.25}\smash{\begin{tabular}[t]{l}{\small t = 1 hour}\end{tabular}}}}%
    \put(0.64329316,0.70925469){\makebox(0,0)[lt]{\lineheight{1.25}\smash{\begin{tabular}[t]{l}{\small t = 100 hours}\end{tabular}}}}%
    \put(0.12600875,0.70925469){\makebox(0,0)[lt]{\lineheight{1.25}\smash{\begin{tabular}[t]{l}{\small t = 50 hours}\end{tabular}}}}%
    \put(0,0){\includegraphics[width=\unitlength,page=2]{gg_LT.pdf}}%
    \put(0.64329316,0.32557895){\makebox(0,0)[lt]{\lineheight{1.25}\smash{\begin{tabular}[t]{l}{\small t = 1000 hours}\end{tabular}}}}%
    \put(0.12600875,0.32557895){\makebox(0,0)[lt]{\lineheight{1.25}\smash{\begin{tabular}[t]{l}{\small t = 500 hours}\end{tabular}}}}%
    \put(0,0){\includegraphics[width=\unitlength,page=3]{gg_LT.pdf}}%
  \end{picture}%
\endgroup%

%% file: Figures/gg_ML.pdf_tex
\begingroup%
  \makeatletter%
  \providecommand\color[2][]{%
    \errmessage{(Inkscape) Color is used for the text in Inkscape, but the package 'color.sty' is not loaded}%
    \renewcommand\color[2][]{}%
  }%
  \providecommand\transparent[1]{%
    \errmessage{(Inkscape) Transparency is used (non-zero) for the text in Inkscape, but the package 'transparent.sty' is not loaded}%
    \renewcommand\transparent[1]{}%
  }%
  \providecommand\rotatebox[2]{#2}%
  \newcommand*\fsize{\dimexpr\f@size pt\relax}%
  \newcommand*\lineheight[1]{\fontsize{\fsize}{#1\fsize}\selectfont}%
  \ifx\svgwidth\undefined%
    \setlength{\unitlength}{345.16403515bp}%
    \ifx\svgscale\undefined%
      \relax%
    \else%
      \setlength{\unitlength}{\unitlength * \real{\svgscale}}%
    \fi%
  \else%
    \setlength{\unitlength}{\svgwidth}%
  \fi%
  \global\let\svgwidth\undefined%
  \global\let\svgscale\undefined%
  \makeatother%
  \begin{picture}(1,1.11006014)%
    \lineheight{1}%
    \setlength\tabcolsep{0pt}%
    \put(0,0){\includegraphics[width=\unitlength,page=1]{gg_ML.pdf}}%
    \put(0.12707173,1.09074513){\makebox(0,0)[lt]{\lineheight{1.25}\smash{\begin{tabular}[t]{l}{\small t = 0 hours}\end{tabular}}}}%
    \put(0.64372697,1.09266214){\makebox(0,0)[lt]{\lineheight{1.25}\smash{\begin{tabular}[t]{l}{\small t = 1 hour}\end{tabular}}}}%
    \put(0.64372697,0.70846514){\makebox(0,0)[lt]{\lineheight{1.25}\smash{\begin{tabular}[t]{l}{\small t = 100 hours}\end{tabular}}}}%
    \put(0.12707173,0.70846514){\makebox(0,0)[lt]{\lineheight{1.25}\smash{\begin{tabular}[t]{l}{\small t = 50 hours}\end{tabular}}}}%
    \put(0.64372697,0.32525611){\makebox(0,0)[lt]{\lineheight{1.25}\smash{\begin{tabular}[t]{l}{\small t = 1000 hours}\end{tabular}}}}%
    \put(0.12707173,0.32525611){\makebox(0,0)[lt]{\lineheight{1.25}\smash{\begin{tabular}[t]{l}{\small t = 500 hours}\end{tabular}}}}%
    \put(0,0){\includegraphics[width=\unitlength,page=2]{gg_ML.pdf}}%
  \end{picture}%
\endgroup%